\documentclass[final,1p,times,authoryear]{elsarticle}

\usepackage{amsmath,amssymb,amsfonts}

\usepackage[english]{babel}
\usepackage{booktabs} 
\usepackage{multirow} 
\usepackage{arydshln} 
\usepackage[hyphens]{url}
\usepackage{textpos} 
\usepackage[acronym]{glossaries}
\usepackage{pdflscape}
\usepackage{mhchem}

\usepackage{tikz} 
\usetikzlibrary{arrows, arrows.meta, plotmarks, positioning, shapes.misc, shapes.geometric, quotes, calc}

\newacronym{sar}{SAR}{Synthetic Aperture Radar}
\newacronym{ari}{ARI}{Anthocyanin Reflectance Index }
\newacronym{mari}{MARI}{Modified Anthocyanin Reflectance Index }
\newacronym{muwi}{MuWi}{Multi-SpectralWater Index}
\newacronym{ndmi}{NDMI}{Normalized Difference Moisture Index}
\newacronym{mndwi}{MNDWI}{Modified Normalized Difference Water Index}
\newacronym{swir}{SWIR}{Short-Wave Infrared}
\newacronym{nbrt1}{NBRT1}{Normalized Burn Ratio Thermal Index}
\newacronym{nbr}{NBR}{Normalized Burn Ratio Index}
\newacronym{bai}{BAI}{Burning Area Index}
\newacronym{asi}{ASI}{Artificial Surface Index}
\newacronym{ri}{RI}{Road Index}
\newacronym{rei}{REI}{Road Extraction Index}
\newacronym{fvc}{FVC}{Fraction of Vegetation Cover}
\newacronym{sza}{SZA}{Solar Zenith Angle}
\newacronym{eu}{EU}{European Union}
\newacronym{jrc}{JRC}{Joint Research Centre}
\newacronym{effis}{EFFIS}{European Forest Fire Information System}
\newacronym{lai}{LAI}{Leaf Area Index}
\newacronym{agb}{AGB}{Above-Ground Biomass}
\newacronym{nir}{NIR}{near-infrared}
\newacronym{lidar}{LIDAR}{Light Detection and Ranging}
\newacronym{mse}{MSE}{Mean Squared Error}
\newacronym{cnn}{CNN}{Convolutional Neural Network}
\newacronym{fnc}{FNC}{Fully Convolution Network}
\newacronym{dem}{DEM}{Digital Elevation Model}
\newacronym{lstm}{LSTM}{Long Short-Term Memory}
\newacronym{vi}{VI}{Vegetation Index}
\newacronym{vwc}{VWC}{Vegetation Water Content}
\newacronym{cir}{CIR}{Color-Infrared}
\newacronym{ndvi}{NDVI}{Normalized Difference Vegetation Index}
\newacronym{ndre}{NDRE}{Normalized Difference Red Edge Index}
\newacronym{ndwi}{NDWI}{Normalized Difference Water Index}
\newacronym{rvi}{RVI}{Radio Vegetation Index}
\newacronym{savi2}{SAVI2-N}{Soil-Adjusted Vegetation Index 2}
\newacronym{ndvi-n}{NDVI-N}{Normalized Difference Vegetation Index (Narrow)}
\newacronym{savi2-n}{SAVI2-N}{Soil-Adjusted Vegetation Index (Narrow)}
\newacronym{viupd}{VIUPD}{Vegetation Index Based on Universal Pattern Decomposition Method}
\newacronym{updm}{UPDM}{Universal Pattern Decomposition Method}
\newacronym{nhvi2}{NHVI2}{Normalized Hotspot-Signature Vegetation Index 2}
\newacronym{hsvi}{HSVI}{Hotspot-Signature Soil-adjusted Vegetation Index}
\newacronym{hevi2}{HEVI2}{Hotspot-Signature 2-Band Enhanced Vegetation Index}
\newacronym{ndhd}{NDHD}{Normalized Difference Between Hotspot and Darkspot}
\newacronym{dart}{DART}{Discrete Anisotropic Radiative Transfer}
\newacronym{boreas}{BOREAS}{Boreal Ecosystem-Atmosphere Study}
\newacronym{tavi}{TAVI}{Topography-Adjusted Vegetation Index}
\newacronym{eo}{EO}{Earth Observation}
\newacronym{sfide}{SFIDE}{Satellite FIre DEtection}
\newacronym{landfire}{LANDFIRE}{Landscape Fire and Resource Management Planning Tools}
\newacronym{abi}{ABI}{Advanced Baseline Imager}
\newacronym{noaa}{NOAA}{National Oceanic and Atmospheric Administration}
\newacronym{leo}{LEO}{Low Earth Orbit}
\newacronym{ros}{ROS}{Rate of Spread}
\newacronym{ca}{CA}{Cellular Automata}
\newacronym{cgfsm}{CGFSM}{CSIRO Grassland Fire Spread Model}
\newacronym{dvi}{DVI}{Difference Vegetation Index}
\newacronym{idvi}{IDVI}{Inverted Difference Vegetation Index}
\newacronym{rdvi}{RDVI}{Renormalized Difference Vegetation Index}
\newacronym{revi}{REVI}{Red-Edge Vegetation Index}
\newacronym{gndvi}{GNDVI}{Green Normalized Difference Vegetation Index}
\newacronym{savi}{SAVI}{Soil-Adjusted Vegetation Index}
\newacronym{msavi}{MSAVI}{Modified Soil-Adjusted Vegetation Index}
\newacronym{osavi}{OSAVI}{Optimized Soil-Adjusted Vegetation Index}
\newacronym{sr}{SR}{Simple Ratio Index}
\newacronym{msr}{MSR}{Modified Simple Ratio Index}
\newacronym{evi}{EVI}{Enhanced Vegetation Index}
\newacronym{evi2}{EVI2}{Enhanced Vegetation Index 2}
\newacronym{mtvi2}{MTVI2}{Modified Triangular Vegetation Index}
\newacronym{mcari}{MCARI}{Modified Chlorophyll Absorption in Reflectance Index}
\newacronym{tcari}{TCARI}{Transformed Chlorophyll Absorption in Reflectance Index}
\newacronym{arvi}{ARVI}{Atmospherically Resistant Vegetation Index}
\newacronym{fvi}{FVI}{Fusion Vegetation Index}
\newacronym{pri}{PRI}{Photochemical Reflectance Index}
\newacronym{uav}{UAV}{Unmanned Aerial Vehicle}
\newacronym{gee}{GEE}{Google Earth Engine}
\newacronym{gari}{GARI}{Green Atmospherically Resistant Vegetation Index}
\newacronym{gdvi}{GDVI}{Green Difference Vegetation Index}
\newacronym{grvi}{GRVI}{Green Ratio Vegetation Index}
\newacronym{mbi}{MBI}{Modified Bare Soil Index}
\newacronym{embi}{EMBI}{Enhanced Modified Bare Soil Index}
\newacronym{af}{AF}{Artificial Surface Factor}
\newacronym{vsf}{VSF}{Vegetation Suppressing Factor}
\newacronym{ssf}{SSF}{Soil Suppressing Factol}
\newacronym{mf}{MF}{Modulation Factor}
\newacronym{miou}{MIoU}{Mean Intersection over Union}
\newacronym{rmse}{RMSE}{Root Mean Square Error}
\newacronym{jmd}{JMD}{Jeffries–Matusita Distance}
\newacronym{td}{TD}{Transformed Divergence}
\newacronym{sdi}{SDI}{Spectral Discrimination Index}
\newacronym{bais2}{BAIS2}{Burned Area Index for Sentinel 2}
\newacronym{msi}{MSI}{Multi-Spectral Instrument}
\newacronym{abai}{ABAI}{Analytic Burned Area Index}
\newacronym{tssa}{TSSA}{Time-Series Segment Analysis}

\makeatletter
\def\ps@pprintTitle{   \let\@oddhead\@empty
   \let\@evenhead\@empty
   \def\@oddfoot{\reset@font\hfil\thepage\hfil}
   \let\@evenfoot\@oddfoot
}
\makeatother

\begin{document}

\begin{frontmatter}

\title{Multispectral Indices for Wildfire Management}

\author[label1]{Afonso Fernandes-Oliveira}
\author[label2]{João P. Matos-Carvalho}
\author[label3,label4]{Filipe Moutinho}
\author[label1,label5]{Nuno Fachada}

\affiliation[label1]{organization={Copelabs, Lusófona University},
            addressline={Campo Grande, 376},
            city={Lisboa},
            postcode={1749-024},
            country={Portugal}}

\affiliation[label2]{organization={LASIGE, Departamento de Informática, Faculdade de Ciências, Universidade de Lisboa},
    city={Lisboa},
    postcode={1749-016},
    country={Portugal}
}

\affiliation[label3]{organization={Center of Technology and Systems (UNINOVA-CTS) and Associated Lab of Intelligent Systems (LASI)},
            city={Caparica},
            postcode={2829-516},
            country={Portugal}}

\affiliation[label4]{organization={NOVA School of Science and Technology, NOVA University Lisbon},
            city={Caparica},
            postcode={2829-516},
            country={Portugal}}

\affiliation[label5]{organization={INESC INOV},
            city={Lisboa},
            postcode={1000-029},
            country={Portugal}}

\begin{abstract}
The increasing frequency and severity of wildfires necessitates advanced methods for effective surveillance and management, as traditional ground-based techniques often struggle to adapt to rapidly changing fire behavior and environmental conditions. This study investigates the use of multispectral aerial and satellite imagery for wildfire management through an assessment of current literature and two practical case studies. We evaluate several multispectral indices for their ability to extract environmental features critical for analyzing wildfire behavior, including vegetation, water bodies, and artificial structures. Our results highlight NVDI for vegetation, MNDWI for water features, and MSR for artificial structures as particularly effective for segmentation and feature extraction. The application of these indices enhances wildfire data processing and supports improved monitoring, risk assessment, and response strategies, demonstrating the potential of multispectral imagery to complement traditional wildfire monitoring and management approaches.
\end{abstract}

\begin{keyword}
wildfire \sep
multispectral imaging \sep
multispectral indices \sep
remote sensing \sep
geoscience
\end{keyword}

\end{frontmatter}

\begin{textblock*}{190mm}(-3cm,-17.18cm)
    {\noindent \footnotesize \color{black!90}The peer-reviewed version of this paper is
    published in Frontiers in Remote Sensing at \url{https://doi.org/10.3389/frsen.2026.1807451}.
    This version is typeset by the authors and differs only in pagination and
    typographical detail.}
\end{textblock*}

\section{Introduction}
\label{sec:introduction}

The effective monitoring and maintenance of forests and other vegetated regions are crucial for mitigating the rising risks associated with extreme natural events~\citep{korena-2023}. The growing frequency of extreme heatwaves, lightning storms, and droughts has revealed their destructive potential, especially in terms of escalating the incidence of severe wildfires~\citep{nifc-2022, ec-jrc-2020}. These events highlight the urgent need for strategies that improve the preparedness and response capabilities of civil authorities~\citep{fachada-2022}.

Traditional wildfire monitoring approaches---such as ground-based observations and manual data collection---are often slow to deploy and lack the spatial and temporal resolution required to capture rapid changes in fire behavior and environmental conditions~\citep{filkov2018improving, barmpoutis2020review}. Moreover, many existing predictive models rely on outdated assumptions, with limited integration of recent climate datasets and evolving vegetation dynamics, leading to incomplete or unreliable risk assessments~\citep{xu2024wildfire}.

Advances in remote sensing and satellite-based Earth observation have substantially improved wildfire monitoring, providing frequent, large-scale, and repeatable measurements of land surface conditions. Multispectral imagery, in particular, enables the derivation of spectral indices that characterize vegetation health, fuel moisture, water availability, burn severity, and proximity to human infrastructure---key factors influencing wildfire ignition, spread, and impact~\citep{roy2014landsat}. Despite the increasing availability of remote sensing data for wildfire analysis, existing reviews of remote sensing applications for wildfire management remain relatively limited, particularly with respect to the systematic comparison of multispectral indices and their demonstrated applicability in real-world case studies~\citep{chen2024remote, guiop2025remote}. While several studies provide broader overviews of wildfire monitoring using remote sensing technologies~\citep{chen2024remote, guiop2025remote, chuvieco2019historical} and others review vegetation indices as a general remote sensing topic~\citep{bannari1995review, xue2017significant}, comparatively few works attempt to organize and evaluate multispectral indices according to their relevance for different wildfire-related environmental features, such as vegetation, soil conditions, water bodies, artificial structures, and burned areas.

This paper addresses this gap through a review of multispectral aerial and satellite imagery for wildfire management, focusing on the use and relevance of selected multispectral indices. Specifically, we examine the most effective indices for extracting and segmenting key wildfire-related features, including combustible vegetation, water bodies, soil characteristics, burned areas, and artificial structures.
In addition, we assess commonly used evaluation metrics and discuss the strengths, limitations, and practical applicability of these indices across different wildfire management scenarios.
Drawing on current knowledge and best practices, this work aims to support researchers and practitioners---such as fire ecologists, remote sensing specialists, forest managers, and emergency planners---in enhancing wildfire prevention, monitoring, and response strategies.

The remainder of this paper is organized as follows. Section~\ref{sec:methodology} details the review methodology, outlining the search strategy and the selection criteria employed for the literature. Section~\ref{sub:multispectral} introduces multispectral imaging, providing an overview of the indices reviewed and discussing their relevance and potential.
Section~\ref{sub:metrics} highlights key metrics for evaluating multispectral indices.
Section~\ref{sub:extraction} focuses on feature extraction, including vegetation (Subsection~\ref{sub:veg}) and soil (Subsection~\ref{sub:soil}) attributes, mapping water features (Subsection~\ref{sub:water}) and artificial structures (Subsection~\ref{sub:artificial}), estimating post-fire burnt areas (Subsection~\ref{sub:burnt}), and analyzing the spatiotemporal and multiscale dynamics of multispectral indices (Subsection~\ref{sub:spatiotemporal-dynamics}). Section~\ref{sec:case-studies} presents two practical case studies illustrating the application of these indices. Building on the reviewed literature and case-study results, Section~\ref{sub:recommendations} offers recommendations for applying multispectral indices across different wildfire management scenarios. Section~\ref{sub:limitations} outlines the constraints and limitations regarding the applicability of this research, after which the paper concludes with Section~\ref{sub:conclusions}, summarizing the key findings of this study.

\section{Review Methodology}
\label{sec:methodology}

This work follows a structured narrative review approach aimed at identifying multispectral indices relevant to wildfire monitoring and management. The literature was surveyed through searches in major scientific databases, including Google Scholar, Scopus, and IEEE Xplore, covering publications from the early development of spectral indices in remote sensing to recent work in wildfire-related applications.

Search queries combined terms such as ``multispectral index'', ``vegetation index'', ``remote sensing'', ``burned area'', ``water index'', ``soil index'', and ``wildfire monitoring''. Additional relevant studies were identified through backward and forward citation tracking of key papers.

Indices were included when they (i) rely on multispectral reflectance bands commonly available in satellite or aerial sensors, and (ii) are reported in the literature as useful for extracting environmental features relevant to wildfire management, such as vegetation, soil, water bodies, artificial structures, or burnt areas. The resulting corpus was iteratively refined during the study to ensure coverage of both widely used indices and representative variations proposed in the literature.

This survey does not aim to be exhaustive, but rather to provide a representative overview of indices that have demonstrated relevance in wildfire-related remote sensing tasks.

\section{Multispectral Imaging}
\label{sub:multispectral}

Multispectral imaging is a technique that captures image data across multiple wavelengths in the electromagnetic spectrum. By dividing the spectrum into discrete bands, it enables the detection and analysis of specific spectral signatures emitted or reflected by various land cover types. Although the precise spectral range of each band varies by instrument, band names and spectral intervals typically follow standardized conventions.

Multispectral imaging captures a limited number of spectral bands, usually ranging from 3 to 10, making data acquisition and processing fast and cost-effective. This simplicity and efficiency make it well-suited for broad applications such as agriculture, forestry, and land cover mapping, where general spectral differences are sufficient. However, its relatively low spectral resolution may overlook subtle spectral features, limiting its ability to distinguish between materials with similar spectral properties. In contrast, hyperspectral imaging captures hundreds of narrow, contiguous spectral bands~\citep{chang2013hyperspectral}, providing much higher resolution that allows for precise identification and differentiation of similar materials. This capability is ideal for more complex applications, such as mineral exploration, environmental monitoring, and medical diagnostics, where detailed spectral information is crucial. However, the increased data volume in hyperspectral imaging leads to slower data acquisition and processing, and higher storage costs. Additionally, managing and analyzing hyperspectral data requires significant computational and technical resources, further increasing its complexity and expense. Overall, multispectral imaging is more suitable for wildfire management than hyperspectral imaging because it provides essential data across a few broad spectral bands. This makes it faster, less data-intensive, and more practical for real-time monitoring of large areas.

\begin{figure}
    \centering
    \includegraphics[width=1\linewidth]{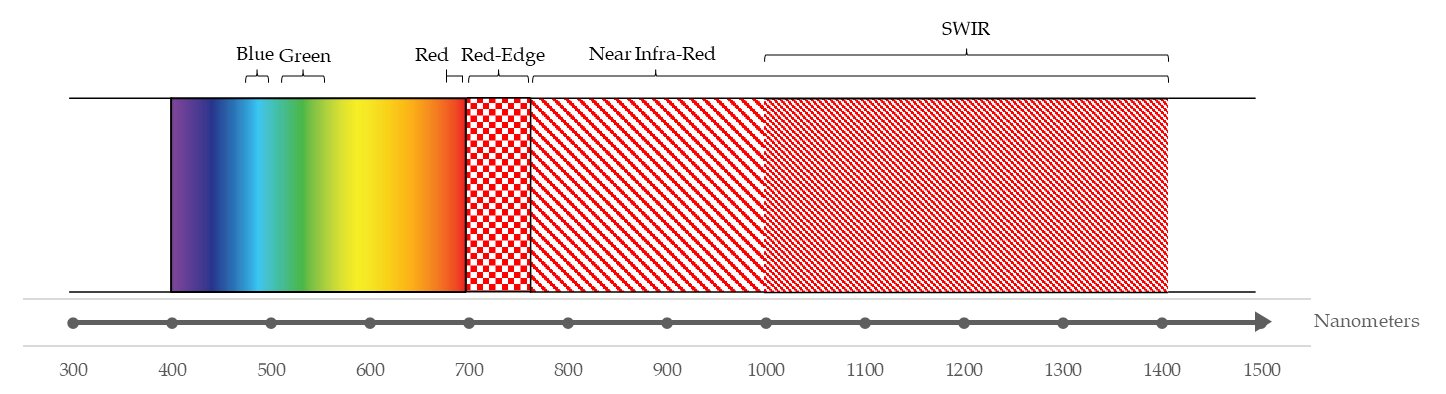}
    \caption{Electromagnetic spectrum with approximate location of relevant bands.}
    \label{fig:em}
\end{figure}

Within multispectral imaging, the blue, green, red, red-edge, and \gls{nir} spectral bands stand out in terms of cost and availability, making them suitable not only for wildfire management \citep{szpakowski2019review}, but also for a wide range of applications, such as fire ecology \citep{perez2020evaluation}, precision agriculture \citep{li2025improved}, and land cover assessment \citep{tian2023simultaneous}. Figure~\ref{fig:em} provides a visual illustration of these and other relevant bands, such as those in the \gls{swir} spectrum, along with their approximate ranges.

Depending on the goal, these bands can be utilized in various ways. For example, the red-edge (700 to 750 nm) \citep{ali-2018} and \gls{nir} (750 to 1400 nm) \citep{jia-2021, veraverbeke-2012} bands can provide valuable insights into the structural and physiological characteristics of vegetation. These are particularly useful for detecting subtle changes in plant health and identifying stress factors.

While single-band analysis may be sufficient for simpler cases, many applications require more advanced approaches, such as generating and analyzing \gls{cir} imagery or computing multispectral indices. \gls{cir} imagery converts \gls{nir} data into the visible spectrum by combining green, red, and \gls{nir} bands. This technique provides a valuable resource that can be analyzed independently or subjected to further processing \citep{foschi-2004, polewski-2021}. \gls{cir} imagery enables more precise assessments of vegetation distribution, density, and health, while also distinguishing between vegetation, soils, water bodies, and certain man-made structures \citep{aronoff-2005}.

In \gls{cir} imagery, healthy, dense vegetation appears in bright red, gradually transitioning to shades of pink as vegetation health declines. Dead vegetation is represented by hues of green, cyan, and tan. Bare soil colors vary depending on composition: clay soils appear as darker shades of tan and cyan, while sandy soils are lighter in tone, ranging from tan to gray or even white. The natural appearance of soil in \gls{cir} imagery can be influenced by factors such as moisture content and organic material concentration \citep{aronoff-2005}. Water bodies are depicted in various shades of blue and black, reflecting their clarity, except in the case of shallow streams, where the appearance directly corresponds to the soil composition of the stream bed. The appearance of man-made structures varies based on their material composition. For instance, gravel roads tend to be lighter, while asphalt roads appear dark blue or black \citep{aronoff-2005}.

Multispectral indices offer a more refined approach to extract insights from multispectral data. These indices enhance specific image bands to create composite images that highlight particular features, thereby improving both the quantity and quality of the data. Various classes of indices have been developed for diverse applications, ranging from vegetation-focused indices to specialized indices for soil content assessment \citep{ramos-2020} and oil slick detection \citep{dong-2021}. Nonetheless, despite rapid advancements in this field, some indices have maintained their foundational importance and continue to be highly relevant.

Table~\ref{tab:multispectral-indices} provides an overview of the indices discussed in this work, categorized by their general formulaic configuration and function, along with the respective formulas and references to the studies that introduced them. These indices will be explored in greater detail throughout the paper. Additionally, Suplementary Table~\ref{tab:acronym-indices} (in Appendix~\ref{sub:annex-acronyms}) presents an alphabetically organized list of acronyms and full names for all indices discussed.

\clearpage
\begin{landscape}

    \begin{table}
\begin{center}
\caption{Multispectral indices grouped by formulaic configuration and function, their respective formula, and a reference to the study that introduced them. \MakeUppercase{\romannumeral 1} - Simple greenness indicators; \MakeUppercase{\romannumeral 2} - Anthocyanin reflectance indices; \MakeUppercase{\romannumeral 3} - Enhanced vegetation indices; \MakeUppercase{\romannumeral 4} - Soil-adjusted vegetation indices; \MakeUppercase{\romannumeral 5} - Modified bare soil indices; \MakeUppercase{\romannumeral 6} - Terrain adjusted vegetation index; \MakeUppercase{\romannumeral 7} - \acrshort{ndhd} composites; \MakeUppercase{\romannumeral 8} - \gls{viupd}; \MakeUppercase{\romannumeral 9} - Water/moisture extraction indices; \MakeUppercase{\romannumeral 10} - Burnt area extraction indices; \MakeUppercase{\romannumeral 11} - Artificial surface index and components; \MakeUppercase{\romannumeral 12} - Road extraction indices;\label{tab:multispectral-indices}}
\resizebox{\linewidth}{!}{
\begin{tabular}{clllclll}
    \toprule
    Group & Index & Formula & Ref. & Group & Index & Formula & Ref.\\
    \cmidrule(rl){1-4}\cmidrule(rl){5-8}
    \MakeUppercase{\romannumeral 1} & SR
        & $\frac{\rho_{\text{NIR}}}{\rho_{\text{R}}}$
        & \citep{birth-1968}

    & \MakeUppercase{\romannumeral 6} & TAVI
        & $\frac{\rho_{\text{NIR}}+f(\Delta)\times (M_{\text{R}}-\rho_{\text{R}})}{\rho_{\text{R}}}$
        & \citep{jiang-2011}\\

    \cdashline{5-8}[1pt/1pt]

    & NDVI
        & $\frac{\rho_{\text{NIR}}-\rho_{\text{R}}}{\rho_{\text{NIR}}+\rho_{\text{R}}}$
        & \citep{rouse-1974}

    & \MakeUppercase{\romannumeral 7} & NDHD
        & $\frac{\rho_{\text{HS}}-\rho_{\text{DS}}}{\rho_{\text{HS}}+\rho_{\text{DS}}}$
        & \citep{leblanc-2001} \\

    & DVI
       & $\rho_{\text{NIR}}-\rho_{\text{R}}$
       & \citep{tucker-1979}

    & & NHVI2
        & $\text{NDVI}\times\text{NDHD}$
        & \citep{zhen-2020}\\
    & RDVI
        & $\frac{\rho_{\text{NIR}}-\rho_{\text{R}}}{\sqrt{\rho_{\text{NIR}}+\rho_{\text{R}}}}$
        & \citep{roujean-1995}

    & & HSVI
        & $\text{SAVI}\times\text{NDHD}$
        & \citep{zhen-2020}\\

    & MSR
        & $\frac{\frac{\rho_{\text{NIR}}}{\rho_{\text{R}}}-1}{\sqrt{\frac{\rho_{\text{NIR}}}{\rho_{\text{R}}}+1}}$
        & \citep{chen-1996}

    & & HEVI2
        & $\text{EVI2}\times\text{NDHD}$
        & \citep{zhen-2020}\\

    \cdashline{5-8}[1pt/1pt]

    & GNDVI
        & $\frac{\rho_{\text{NIR}}-\rho_{\text{G}}}{\rho_{\text{NIR}}+\rho_{\text{G}}}$
        & \citep{gitelson-1996}

    & \MakeUppercase{\romannumeral 8} & VIUPD
        & $\frac{C_\text{V}-0.12\times C_\text{S}-C_\text{4}}{C_\text{W}+C_\text{V}+C_\text{S}}$
        & \citep{zhang-2007}\\

    \cdashline{5-8}[1pt/1pt]

    & GARI
        & $\frac{\rho_{\text{NIR}}-[\rho_{\text{G}}-\gamma(\rho_{\text{B}}-\rho_{\text{R}})]}{\rho_{\text{NIR}}+[\rho_{\text{G}}-\gamma(\rho_{\text{B}}-\rho_{\text{R}})]}$
        & \citep{gitelson-1996}

    & \MakeUppercase{\romannumeral 9} & NDWI
        & $\frac{\rho_{\text{G}}-\rho_{\text{NIR}}}{\rho_{\text{G}}+\rho_{\text{NIR}}}$
        & \citep{gao-1996}\\

    & NDRE
        & $\frac{\rho_{\text{NIR}}-\rho_{\text{RE}}}{\rho_{\text{NIR}}+\rho_{\text{RE}}}$
        & \citep{barnes-2000}

    & & NDMI
        & $\frac{\rho_{\text{NIR}}-\rho_{\text{SWIR}}}{\rho_{\text{NIR}}+\rho_{\text{SWIR}}}$
        & \citep{gao-1996}\\

    & GDVI
        & $\rho_{\text{NIR}}-\rho_{\text{G}}$
        & \citep{sripada-2005}

    & & MNDWI
        & $\frac{\rho_{\text{G}}-\rho_{\text{SWIR}}}{\rho_{\text{G}}+\rho_{\text{SWIR}}}$
        & \citep{xu-2006}\\

    \cdashline{5-8}[1pt/1pt]

    & GRVI
        & $\frac{\rho_{\text{NIR}}}{\rho_{\text{G}}}$
        & \citep{sripada-2006}

    & \MakeUppercase{\romannumeral 10} & NBR
        & $\frac{\rho_{\text{NIR}}-\rho_{\text{SWIR}}}{\rho_{\text{NIR}}+\rho_{\text{SWIR}}}$
        & \citep{lopez-1991}\\

    & IDVI
        & $\frac{1+(\rho_{\text{NIR}}-\rho_{\text{R}})}{1-(\rho_{\text{NIR}}-\rho_{\text{R}})}$
        & \citep{sun-2018}

    & & BAI
        & $\frac{1}{(0.12-\text{R})^2+(0.06-\text{NIR})^2}$
        & \citep{chuvieco-1998}\\

    \cdashline{1-4}[1pt/1pt]

    \MakeUppercase{\romannumeral 2} & ARI
        & $\frac{1}{\rho_{\text{G}}}-\frac{1}{\rho_{\text{RE}}}$
        & \citep{gitelson-2009}

    & & NBRT1
        & $\frac{\rho_{\text{NIR}}-\rho_{\text{SWIR}}(\frac{\rho_{\text{Thermal}}}{1000})}{\rho_{\text{NIR}}+\rho_{\text{SWIR}}(\frac{\rho_{\text{Thermal}}}{1000})}$
        & \citep{holden-2005}\\

    \cdashline{5-8}[1pt/1pt]

    & MARI
        & $\text{ARI}\times\rho_{\text{R}}$
        & \citep{gitelson-2009}

    & \MakeUppercase{\romannumeral 11} & ASI
        & $\phi(\text{AF})\times \phi(\text{VSF})\times \phi(\text{SSF})\times \phi(\text{MF})$
        & \citep{zhao-2022}\\

    \cdashline{1-4}[1pt/1pt]

    \MakeUppercase{\romannumeral 3} & EVI
        & $\frac{2.5\times (\rho_{\text{NIR}}-\rho_{\text{R}})}{\rho_{\text{NIR}}+6\times \rho_{\text{R}}-7.5\times \rho_{\text{B}}+1}$
        & \citep{liu-1995}

    & & ASI (Alt.)
        & $\frac{1}{16}\times (\text{AF} + 1)\times \text{VSF}\times \text{SSF}\times (\text{MF} + 1)$
        & \citep{zhao-2022}\\

    & EVI2
        & $\frac{2.5\times (\rho_{\text{NIR}}-\rho_{\text{R}})}{\rho_{\text{NIR}}+2.4\times \rho_{\text{R}}+1}$
        & \citep{jiang-2006}

    & & AF
        & $\frac{\rho_{\text{NIR}}-\rho_{\text{B}}}{\rho_{\text{NIR}}+\rho_{\text{B}}}$
        & \citep{zhao-2022}\\

    \cdashline{1-4}[1pt/1pt]

    \MakeUppercase{\romannumeral 4} & SAVI
            & $1.5\times \frac{\rho_{\text{NIR}}-\rho_{\text{R}}}{\rho_{\text{NIR}}+\rho_{\text{R}}+0.5}$
            & \citep{huete-1988}

    & & VSF
        & $1-(\text{NDVI}\times \text{MSAVI})$
        & \citep{zhao-2022}\\

    & MSAVI
        & $\frac{2\times \rho_{\text{NIR}}+1-\sqrt{(2\times \rho_{\text{NIR}}+1)^2-8\times (\rho_{\text{NIR}}-\rho_{\text{RED}})}}{2}$
        & \citep{qi-1994}

    & & SSF
        & $1-\text{EMBI}$
        & \citep{zhao-2022}\\

    & OSAVI
        & $1.126\times \frac{\rho_{\text{NIR}}-\rho_{\text{R}}}{\rho_{\text{NIR}}+\rho_{\text{R}}+0.126}$
        & \citep{rendeaux-1996}

    & & MF
        & $\frac{(\rho_{\text{B}}+\rho_{\text{G}})-(\rho_{\text{NIR}}+\rho_{\text{SWIR1}})}{(\rho_{\text{B}}+\rho_{\text{G}})+(\rho_{\text{NIR}}+\rho_{\text{SWIR1}})}$
        & \citep{zhao-2022}\\

    \cdashline{1-4}[1pt/1pt]
    \cdashline{5-8}[1pt/1pt]

    \MakeUppercase{\romannumeral 5} & MBI
        & $\frac{\rho_{\text{SWIR1}}-\rho_{\text{SWIR2}}-\rho_{\text{NIR}}}{\rho_{\text{SWIR1}}+\rho_{\text{SWIR2}}+\rho_{\text{NIR}}}+0.5$
        & \citep{nguyen-2021}

    & \MakeUppercase{\romannumeral 12} & REI
        & $\frac{\rho_{\text{NIR}}-\rho_{\text{B}}}{\rho_{\text{NIR}}+\rho_{\text{B}}\times \rho_{\text{NIR}}}$
        & \citep{shahi-2015}\\

    & EMBI
        & $\frac{\text{MBI}-\text{MNDWI}-0.5}{\text{MBI}+\text{MNDWI}+1.5}$
        & \citep{zhao-2022}

    & & RI
        & $1-\frac{3\times min(\rho_{\text{SWIR}},\rho_{\text{NIR}},\rho_{\text{B}})}{\rho_{\text{SWIR}}+\rho_{\text{NIR}}+\rho_{\text{B}}}$
        & \citep{waqas-2022}\\

    \bottomrule

    \multicolumn{8}{l}{\scriptsize $\qquad\rho_{\text{B}}$, $\rho_{\text{G}}$, $\rho_{\text{R}}$, $\rho_{\text{RE}}$, $\rho_{\text{NIR}}$, $\rho_{\text{SWIR}}$: Reflectance for the Blue, Green, Red, Red-Edge, NIR, and SWIR band reflectances, respectively.}\\
    \multicolumn{8}{l}{\scriptsize $\qquad\rho_{\text{Thermal}}$, $\rho_{\text{HS}}$, $\rho_{\text{DS}}$: Thermal, Hotspot, and Darkspot reflectances, respectively.}\\
    \multicolumn{8}{l}{\scriptsize $\qquad{}C_\text{V}$, $C_\text{S}$, $C_\text{W}$, $C_\text{Y}$: Vegetation, Soil, Water, and Yellow reflectance coefficients, respectively.}\\
    \multicolumn{8}{l}{\scriptsize $\qquad{}f(\Delta)$: Topography adjusting coefficient.}\\
    \multicolumn{8}{l}{\scriptsize $\qquad{}M_{\text{RED}}$: Maximized Red band reflectance }\\
    \multicolumn{8}{l}{\scriptsize $\qquad\phi(\#)$: Min–Max normalization function based on the entire image.}\\

    \end{tabular}}
\end{center}
\end{table}
\end{landscape}
\clearpage

\section{Index Comparability and Metrics}
\label{sub:metrics}

To facilitate the interpretation and comparison of results reported in the literature review, this section provides an overview of the most commonly used performance metrics for evaluating multispectral indices and related methods. The aim is to establish a common reference framework that supports a consistent and informed understanding of the metrics employed across different studies.

Studies in the existing literature employ various metrics to assess performance, often influenced by the chosen methodology or the author's preferences. One commonly used metric is \textit{accuracy}, which measures how closely a result aligns with the true or expected value. Higher accuracy indicates more correct and reliable predictions, measurements, or outcomes. Accuracy is typically expressed as a percentage, with 100\% indicating that the method perfectly describes or predicts the outcomes without any errors or discrepancies. Another important metric is \textit{sensitivity}, which evaluates how variations in a particular variable or condition influence the outcome. Sensitivity is crucial for assessing a method's robustness and reliability under varying circumstances. In statistical contexts, accuracy and sensitivity are commonly defined using Equations~\ref{eq:accu},~\ref{eq:sens}:

\begin{equation}
\text{Accuracy} = \frac{\text{TP} + \text{TN}}{\text{TP} + \text{TN} + \text{FP} + \text{FN}}
\label{eq:accu}
\end{equation}
\begin{equation}
\text{Sensitivity} = \frac{\text{TP}}{\text{TP} + \text{FN}}
\label{eq:sens}
\end{equation}

\noindent where \textit{TP} denotes true positives, \textit{TN} refers to true negatives, \textit{FP} represents false positives, and \textit{FN} corresponds to false negatives. For example, in the context of pixel-wise image analysis, these metrics could evaluate the classification of individual pixels. A \textit{TP} occurs when a pixel that genuinely belongs to the target class is correctly identified. A \textit{TN} is when a pixel that does not belong to the target class is accurately classified as non-target. A \textit{FP} arises when a pixel that does not belong to the target class is incorrectly classified as part of it, leading to over-detection or false alarms. Conversely, a \textit{FN} occurs when a pixel that belongs to the target class is misclassified as non-target, resulting in an omission or missed detection.

However, beyond accuracy and sensitivity, a range of additional metrics are frequently reported in the literature for comparing multispectral indices. Table~\ref{tab:metrics} summarizes the evaluation metrics most frequently encountered, along with their typical ranges and general interpretation.

\begin{table}
\begin{center}
\caption{Common metrics for comparing multispectral indexes, their expected ranges, and interpretation.\label{tab:metrics}}
\scriptsize
\begin{tabular}{lllc}
    \toprule
    Acronym & Name & Range & Interpretation  \\
    \midrule
    --- &
        Accuracy &  $[0,1]^a$ & $\blacktriangle$ \\
    --- &
        Sensitivity &  $[0,1]^a$ & $\blacktriangle$ \\
    \cmidrule(rl){1-4}
    $R^2$ &
        Coefficient of Determination &  $[0,1]$ & $\blacktriangle$ \\
    RMSE &
        Root-Mean-Square Deviation &  $[0,+\infty]^a$ & $\blacktriangledown$ \\
    MIoU &
         Mean Intersection of Union & $[0,1]$ & $\blacktriangle$ \\
    \cmidrule(rl){1-4}
    t-test &
        Student’s t-Test & $[0,1]^b$ & $\blacktriangledown$ \\
    ANOVA &
        Analysis of Variance & $[0,1]^b$ & $\blacktriangledown$ \\
    MWU &
        Mann-Whitney U Test & $[0,1]^b$ & $\blacktriangledown$ \\
    KW &
        Kruskal-Wallis Test & $[0,1]^b$ & $\blacktriangledown$ \\
    McNemar &
        McNemar Test & $[0,1]^b$ & $\blacktriangledown$ \\
    \cmidrule(rl){1-4}
    JMD &
        Jeffries–Matusita Distance & $[0,\sqrt{2}]^a$ & $\blacktriangle$ \\
    TD &
        Transformed Divergence & $[0,2]^a$ & $\blacktriangle$ \\
    SDI &
        Spectral Discrimination Index & $[-\infty,\infty]^a$ & $\blacktriangle$ \\
    \bottomrule

    \multicolumn{4}{l}{\scriptsize $\blacktriangle$  A higher value is typically better.}\\
    \multicolumn{4}{l}{\scriptsize $\blacktriangledown$ A lower value is typically better.}\\
    \multicolumn{4}{l}{\scriptsize $^a$ Sometimes expressed as a percentage.}\\
    \multicolumn{4}{p{0.9\linewidth}}{\scriptsize $^b$ Each statistical test returns a specific value ($t$-statistic, $F$-statistic, $U$-statistic, $H$-statistic, and $\chi^2$, respectively), along with a $p$-value for interpretation. $p$-values $<0.01$ and $<0.05$ are commonly used thresholds for statistical significance.}\\

\end{tabular}
\end{center}
\end{table}

The $R^2$ value is a metric that reflects how well a model explains the variance in the dependent variable, with its range spanning from 0 to 1. A higher $R^2$ indicates that the model provides a closer fit to the observed data. In contrast, the \gls{rmse} measures the average prediction error of a model. \gls{rmse} values range from 0 to $+\infty$, and lower values indicate that the model's predictions are more aligned with the observed data.

The \gls{miou} is another important metric, particularly used in boundary and segmentation tasks. It evaluates the overlap between predicted regions and the ground truth, with values ranging from 0 to 1, where higher values indicate more precise segmentation performance.

Statistical tests provide a rigorous framework for evaluating differences in data, ensuring that observed variations are not merely the result of random fluctuations. In the context of multispectral indices and remote sensing, the choice of statistical test depends on the characteristics of the data and the specific research objectives, with each test yielding a distinct metric or value. These test usually also generate a $p$-value. A $p$-value expresses the probability of obtaining results as extreme as the observed data, assuming the null hypothesis is true. A small $p$-value (often set at a threshold of 0.05 or 0.01) suggests that the observed result is statistically significant, implying it is unlikely to have occurred under the null hypothesis. In the context of multispectral indices, $p$-values help assess the statistical significance of differences in spectral data across different land cover types, vegetation health, or soil moisture levels. For instance, $p$-values can be used to determine whether observed changes in specific spectral bands (such as SWIR or NIR) between two time periods are statistically significant, supporting conclusions about vegetation stress, fire severity, or water content with more confidence.

In the context of multispectral image analysis, the Student’s t-Test is commonly used to compare the mean values of a spectral index between two independent or paired conditions, assuming normality. It returns a $t$-statistic, which measures the difference between group means relative to variability, and a $p$-value. When comparing more than two groups, Analysis of Variance (ANOVA) determines whether significant differences exist among them. It returns an $F$-statistic, reflecting the ratio of variance between groups to variance within groups. For non-normally distributed data, the Mann-Whitney U Test serves as a non-parametric alternative to the $t$-test, while the Kruskal-Wallis Test extends this approach to multiple groups. These tests return a $U$-statistic or $H$-statistic, respectively, along with a $p$-value, indicating whether distributions significantly differ. Finally, the McNemar Test is applied to paired categorical data, making it useful in classification accuracy assessments and land cover change detection. It returns a $\chi^2$ statistic (or an exact $p$-value in small samples), assessing whether the proportion of changes between categories is statistically significant.

Separability metrics, including the \gls{jmd} \citep{jeffreys1946invariant, matusita1955decision}, \gls{td} \citep{swain1973two}, and \gls{sdi} \citep{kaufman1994detection}, are commonly used to evaluate classification performance. Separability refers to how well different classes within a dataset can be distinguished from one another. While the \gls{jmd} and \gls{td} primarily focus on probabilistic separability, the \gls{sdi} is more tailored to the spectral characteristics, making it especially relevant in remote sensing applications.

The \gls{jmd} quantifies the statistical distance between two probability distributions based on their spatial overlap, as defined in Equation~\ref{eq:jmd-1}:

\begin{equation}
JMD_{ij} = \sqrt{2\left(1 - e^{-B_{ij}}\right)}
\label{eq:jmd-1}
\end{equation}

\noindent where $i$ and $j$ denote the indices of the classes whose probability distributions are being compared. It lies within the interval $[0, \sqrt{2}]$, where $0$ indicates completely overlapping distributions and $\sqrt{2}$ corresponds to fully separable classes.

The \gls{jmd} is derived from the Bhattacharyya distance, $B_{ij}$ \citep{bhattacharyya1943measure}, which can be applied to any pair of probability distributions. However, in remote sensing applications, it is commonly computed under the assumption that class distributions are multivariate normal, which allows a closed-form solution based on class means and covariance matrices, as expressed in Equation~\ref{eq:jmd-2}:

\begin{equation}
B_{ij} = \frac{1}{8}(\mu_i - \mu_j)^T \left(\frac{\Sigma_i + \Sigma_j}{2}\right)^{-1} (\mu_i - \mu_j) + \frac{1}{2} \ln \left( \frac{\det\left(\frac{\Sigma_i + \Sigma_j}{2}\right)}{\sqrt{\det(\Sigma_i)\det(\Sigma_j)}} \right)
\label{eq:jmd-2}
\end{equation}

\noindent where $\mu_i$ and $\mu_j$ are the mean vectors for classes $i$ and $j$, while $\Sigma_i$ and $\Sigma_j$ denote their respective covariance matrices. The term $\det(\cdot)$ represents the matrix determinant.

The \gls{td} scales the divergence between distributions into a bounded range. It is defined as functional transformation of the divergence $D_{ij}$ \citep{swain1973pattern}, as described in Equations~\ref{eq:td-1},~\ref{eq:td-2}:

\begin{equation}
TD_{ij} = 2 \times \left( 1 - e^{-\frac{D_{ij}}{8}} \right)
\label{eq:td-1}
\end{equation}

\begin{equation}
D_{ij} = \frac{1}{2} \text{tr} \left[ (\Sigma_i - \Sigma_j)(\Sigma_j^{-1} - \Sigma_i^{-1}) \right] + \frac{1}{2} \text{tr} \left[ (\Sigma_i^{-1} + \Sigma_j^{-1})(\mu_i - \mu_j)(\mu_i - \mu_j)^T \right]
\label{eq:td-2}
\end{equation}

\noindent where $\text{tr}(\cdot)$ denotes the matrix trace, while $\mu$ and $\Sigma$ represent the mean vectors and covariance matrices for the respective classes $i$ and $j$. The metric is bounded within the interval $[0, 2]$, where $TD_{ij} < 1.0$ denotes poor separability, $1.0 \leq TD_{ij} \leq 1.9$ denotes moderate separability, and $TD_{ij} > 1.9$ denotes very good separability.

The \gls{sdi} evaluates class separability as the ratio between the difference in spectral means and the sum of their standard deviations, as expressed in Equation~\ref{eq:sdi}:

\begin{equation}
SDI_{ij} = \frac{|\mu_i - \mu_j|}{\sigma_i + \sigma_j}
\label{eq:sdi}
\end{equation}

\noindent where $\mu_i$ and $\mu_j$ are the mean spectral reflectance values for classes $i$ and $j$, with $\sigma_i$ and $\sigma_j$ as their respective standard deviations. In practice, class separability is evaluated using the absolute value $\left|SDI\right|$, where values close to zero indicate poor discrimination between classes and higher values indicate greater separability.

\section{Feature Extraction}
\label{sub:extraction}

Multispectral indices are essential tools in remote sensing for feature extraction, as they enable the identification and analysis of specific surface characteristics. These characteristics serve as key data inputs for wildfire prediction algorithms, providing critical information for accurate forecasting. Four common characteristics have been identified as useful for such algorithms  \citep{sullivan2009wildlanda,sullivan2009wildlandb,sullivan2009wildlandc}: the distribution and composition of vegetation (Subsection~\ref{sub:veg}) and soil (Subsection~\ref{sub:soil}), which typically constitute the primary sources of fuel; the arrangement of water features (Subsection~\ref{sub:water}), which can serve as natural fire barriers; the placement of man-made structures (Subsection~\ref{sub:artificial}), which can affect the fire's spread and must thus be considered when formulating wildfire management plans. Post-fire measures are also highly relevant. In particular, the assessment of burnt areas is instrumental in damage analysis, rehabilitation, and reforestation efforts. Consequently, this factor is examined in Subsection~\ref{sub:burnt}. Finally, we examine the distinctions, spatial and temporal dynamics of \textit{static} and \textit{spatiotemporal} analysis approaches, as detailed in Subsection~\ref{sub:spatiotemporal-dynamics}.

Figure~\ref{fig:feature-extraction} depicts an example process flow for extracting wildfire-related insights using multispectral indices. These insights help understand fire growth and movement patterns, which in turn can inform strategies for land management, including optimal locations for firebreaks,  prescribed burns, fuel reduction programs, and evacuation plans.

\tikzstyle{arrow} = [thick,->,>=stealth]
\tikzstyle{central} = [rectangle, rounded corners, minimum width=0.5cm, minimum height=0.5cm,text centered, draw=black]
\tikzstyle{normal} = [rectangle, minimum width=0.5cm, minimum height=0.5cm,text centered, draw=black]

\begin{figure}
    \centering
    \begin{tikzpicture}[node distance=2.5cm, every node/.style={font=\footnotesize}]

    \node (imaging) [normal] {Multispectral Imaging};

    \node (indices) [normal, below of=imaging] {Multispectral Indices};

    \node (processing) [normal, below of=indices] {Processing
        \begin{minipage}{3.5cm}
            \begin{itemize}
                \item Segmentation
                \item Classification
                \item Analysis
            \end{itemize}
        \end{minipage}};

    \node (fuel) [normal, below of=processing] {Fuel Maps
        \begin{minipage}{3cm}
            \begin{itemize}
                \item Distribution
                \item Composition
                \item Density
            \end{itemize}
        \end{minipage}};

    \node (prediction) [normal, below of=fuel] {Wildfire Prediction};

    \node (manag) [normal, below of=prediction, yshift=-0.75cm] {Improved Fire Management};
    \node (risk) [normal, left of=manag, yshift=0.75cm] {Risk Assessment};
    \node (analys) [normal, right of=manag, yshift=0.75cm] {Post-Fire Analysis};

    \draw [arrow] (imaging) -- (indices) node[midway, right] {Bands};;
    \draw [arrow] (indices) -- (processing);
    \draw [arrow] (processing) -- (fuel);
    \draw [arrow] (fuel) -- (prediction);
    \draw [arrow] (prediction) -- (risk);
    \draw [arrow] (prediction) -- (manag);
    \draw [arrow] (prediction) -- (analys);

    \end{tikzpicture}
    \caption{Feature extraction for wildfire prediction using multispectral indices.}
    \label{fig:feature-extraction}
\end{figure}

\subsection{Vegetation}
\label{sub:veg}

Vegetation plays an essential role as a fuel source in wildfire behavior. Its composition, quantity, spatial arrangement, and moisture content directly impacts the intensity, spread, and duration of wildfires. Vegetation indexes are assessed based on their performance in distinguishing different vegetation densities (density performance), resistance to terrain reflection artifacts (topographic resistance), and resistance to atmospheric effects (atmospheric resistance).

The effectiveness of a vegetation index in assessing vegetation coverage relies on its ability to accurately measure density. The most commonly utilized vegetation indexes are those that measure greenness, leveraging the reflectance and absorption properties of chlorophyll pigments to highlight vegetation details, as seen with \gls{ndvi} \citep{rouse-1974}. However, there are also indices that focus on alternative pigments, such as the \gls{ari} \citep{gitelson-2009} and the \gls{mari} \citep{gitelson-2009} indices, which target anthocyanin pigments that typically impart blue, red, or purple hues to vegetation. The effectiveness of an index is closely linked to the density of pigment-rich vegetation in a specific area, making it less effective in regions where vegetation lacks these pigments.

Topographic and atmospheric factors can have a strong impact on the accuracy and reliability of vegetation indices. Topographic considerations include the influence of terrain features, such as slopes and shadows, which can affect sunlight availability and alter reflectance values. Atmospheric factors, on the other hand, are caused by components like aerosols, water vapor, and clouds, which scatter and attenuate solar radiation, leading to errors in reflectance measurements. Techniques such as topographic~\citep{colby1991topographic, gu1998topographic} and atmospheric correction~\citep{berk1999modtran4, vermote2008atmospheric, doxani2018atmospheric} algorithms can mitigate these issues, thereby enhancing the accuracy of vegetation indices. Table~\ref{tab:veg-indices} provides a summary of the most effective indices for extracting vegetation attributes, highlighting their specific strengths.

\begin{table}
\begin{center}
\caption{Multispectral indices for vegetation attribute extraction.\label{tab:veg-indices}}
\small
\begin{tabular}{lcccccl}
    \toprule
    \multirow{2}{*}{Index} & \multicolumn{3}{l}{Density} & \multicolumn{2}{l}{Resistance} & \multirow{2}{*}{HI} \\
    \cmidrule(rl){2-4}\cmidrule(rl){5-6}
    & Low & Med. & High & Top. & Atm. & \\
    \midrule
    NDVI &
        \checkmark & \checkmark & & \checkmark & & \checkmark \\
    SAVI &
        \checkmark & \checkmark & & \checkmark & & \\
    RDVI &
        & \checkmark & \checkmark & & \checkmark & \\
    EVI &
        & \checkmark & \checkmark & \checkmark & \checkmark & \\
    NDWI &
        \checkmark & \checkmark & \checkmark & & & \checkmark \\
    NDRE &
        \checkmark & \checkmark & & \checkmark & & \\
    EVI2 &
        & \checkmark & \checkmark & \checkmark & \checkmark & \\
    VIUPD &
        \checkmark & \checkmark & \checkmark & & & \checkmark \\
    TAVI &
        \checkmark & \checkmark & & \checkmark & & \\
    IDVI &
        & \checkmark & \checkmark & \checkmark & & \\
    NHVI2 &
        \checkmark & \checkmark & \checkmark & & & \\
    HSVI &
        \checkmark & \checkmark & \checkmark & & & \\
    HEVI2 &
        \checkmark & \checkmark & \checkmark & & & \\
    \bottomrule

    \multicolumn{7}{l}{\tiny Top.: Topographic; Atm.: Atmospheric.}\\
    \multicolumn{7}{l}{\tiny HI: Can be used as an health indicator.}\\

    \end{tabular}
\end{center}
\end{table}

When working with multispectral indices, even widely used ones, it is necessary to take into account their limitations. In terms of density performance, \gls{ndvi} has been found to saturate in highly vegetated areas and produce inconsistent results in very arid regions \citep{mummoorthy-2019}---with \gls{ndre} \citep{barnes-2000} being a possible alternative for the latter case \citep{li-2012}. Similarly, \gls{savi} \citep{huete-1988} faces challenges in accurately estimating vegetation in areas with a heterogeneous canopy \citep{zhen-2020}.

Both the \gls{idvi} \citep{sun-2018} and the \gls{viupd} \citep{zhang-2007} indexes have demonstrated their ability to overcome some  of \gls{ndvi}'s limitations. \gls{idvi} shows insensitivity to leaf biochemical parameters and a wider variation range compared to \gls{ndvi}, resulting in more stable results. Similarly, the \gls{viupd} method demonstrates a broader domain range and higher sensitivity to vegetation density, leading to superior performance in areas with high vegetation cover. Additionally, \gls{viupd} also displays higher sensitivity to vegetation health and \ce{CO2} concentration.

The methodology employed by \gls{viupd} is uncommon, as it builds upon the \gls{updm} \citep{zhang-2003}. \gls{viupd} was developed as a redefinition of a revised vegetation index, also rooted in \gls{updm} \citep{daigo-2004}. Despite the apparent complexity involved in calculating its coefficients, these can be computed using a structured three-step process \citep{she-2016}, which greatly simplifies the calculations.

To improve the accuracy and applicability of \gls{savi} in estimating vegetation in a heterogeneous canopy, the \gls{ndhd} \citep{leblanc-2001} composites---\gls{nhvi2} \citep{zhen-2020}, \gls{hsvi} \citep{zhen-2020}, and \gls{hevi2} \citep{zhen-2020}---were developed. These indices are calculated through the multiplication of a base index---\gls{ndvi}, \gls{savi} and \gls{evi2} \citep{zhen-2020}---by the aforementioned \gls{ndhd}. The latter represents the distribution of foliage in a canopy and its derived indices generally performed well when compared to \gls{savi} and other commonly used multispectral vegetation indices, demonstrating higher performance in heterogeneous canopy regions and resistance to soil-noise effects \citep{zhen-2020}.

Zhou et al. \citep{zhou-2019} consider topographic resistance as a notable characteristic shared by ratio indices, such as \gls{ndvi}, \gls{ndwi} \citep{gao-1996}, and \gls{ndre}. These indices---derived from low complexity ratios---have been observed to partially mitigate the influence of topographic variations, especially when compared to non-ratio indices like \gls{evi} \citep{liu-1995,zhou-2019}. The issue of atmospheric resistance can also be minimized by various indices, such as \gls{rdvi}  \citep{roujean-1995}, which offers additional mitigation against specific solar geometry distortions \citep{jingguo-2015}. In addition to \gls{rdvi}, the \gls{evi} and \gls{evi2} indices are also robust in the presence of atmospheric interference.

The \gls{tavi} \citep{jiang-2011} index was developed taking topographic resistance into account. It addresses the high correlation between the solar incidence cosine and conventional vegetation indices that can lead to a loss of accuracy \citep{jiang-2011}. \gls{tavi} introduces two coefficients: \(M_{\text{RED}}\), representing the maximized value of the red waveband, and \(f(\Delta)\), the topography adjusting coefficient.  The \(f(\Delta)\) coefficient, with a value of 2.28, minimizes the mean difference between shadowed and brightly lit inclines. Interestingly, \gls{tavi} relies solely on the red and \gls{nir} bands, without support from additional topographical data.

Several other index qualities warrant consideration, such as utility for specific value estimations or use as alternatives for other indices \citep{gonec-2019}. For instance, when estimating fractional vegetation cover, a \gls{dvi}-based \citep{tucker-1979} model achieved the highest accuracy and stability \citep{yan-2021}. The \gls{ndvi} can also serve as a reliable alternative for estimating the live vegetation fraction, as demonstrated by Laneve et al.~\citep{laneve2024application}. Furthermore, \gls{ndvi} can be a viable alternative to \gls{rvi}, which uses radio band data that tends to be costly and difficult to acquire \citep{gonec-2019}. Finally, \gls{ndwi} exhibits strong accuracy in estimating vegetation water content, and since a plant's water content is often correlated to its condition, \gls{ndwi} can serve as a reliable indicator of vegetation health as well \citep{lu-2011}.

The \gls{lai} \citep{breda2003ground} is a dimensionless measure that represents the amount of leaf surface area relative to a given ground area. Although not a multispectral index, it is commonly used in plant science, forestry, ecology, and agronomy to quantify vegetation density. It is defined as the total leaf area of the canopy divided by the horizontal ground area beneath the canopy, as shown in Equation~\ref{eq:lai}:

\begin{equation}
\text{LAI} = \frac{\text{Total Leaf Area}}{\text{Ground Area}}
\label{eq:lai}
\end{equation}

\gls{lai} has been extensively studied due to its use in ecological and hydrological models \citep{kang-2016, mourad-2020, liu-2021}. Not considering direct and labor-intensive measuring approaches, beyond the scope of this work, \gls{lai} can be estimated through various indirect, multispectral index-based methods, summarized in  Table~\ref{tab:lai}. Methods based on combinations of indices have shown superior accuracy and robustness in \gls{lai} estimation compared to single index models \citep{sun-2018, kalpoma-2019}. Nevertheless, single models for \gls{lai} estimation still exhibit significant correlations with observed values, especially in the \gls{nir}, red, and blue spectral ranges \citep{jingguo-2015}. Regarding solo models, the literature indicates that models based on \gls{ndvi} outperform those relying on simple ratio indices, including \gls{sr} \citep{birth-1968} and \gls{msr} \citep{chen-1996,xie-2014}.

\begin{table}
\begin{center}
\caption{Some \gls{lai} estimation approaches. Adapted from \citep{mourad-2020}.\label{tab:lai}}
\footnotesize
\begin{tabular}{lll}
    \toprule
    Base & Formula & Ref. \\
    \midrule
    \multirow{3}{*}{NDVI} &
        $9.519 \times\text{NDVI}^3-0.1204 \times\text{NDVI}^2+1.236 \times\text{NDVI}-0.257$ & \citep{myneni-1997} \\
    & $4.9\times\text{NDVI}-0.46$ & \citep{johnson-2003} \\
    & $0.0287\times\text{e}^{5.081\times\text{NDVI}}$ & \citep{piqueras-2008} \\
    \midrule
    \multirow{4}{*}{EVI2} &
        $(2.92\times\sqrt{\text{EVI2}}-0.43)^2$ & \multirow{4}{*}{\citep{kang-2016}} \\
    & $(3.126\times\sqrt{\text{EVI2}}-0.58)^2$ &  \\
    & $(5.3\times\sqrt{\text{EVI2}}-1.66)^{\frac{3}{2}}$ & \\
    & $(5.47\times\text{EVI2}^{\frac{3}{5}}-1.03)^{\frac{4}{3}}$ & \\
    \midrule
    \multirow{2}{*}{SAVI} &
        $11\times\text{SAVI}^3$ & \multirow{2}{*}{\citep{pocas-2014}} \\
    & $\frac{-\ln{(\frac{0.69-\text{SAVI}}{0.59})}}{0.91}$ &  \\
    \midrule
    \multirow{2}{*}{NDVI/IDVI} &
        $(1-\alpha)\times LAI_{\text{NDVI}}+\alpha\times LAI_{\text{IDVI}}$ & \multirow{2}{*}{\citep{sun-2018}} \\
    & $\alpha=\frac{1}{1+e^{-k\times (NDVI-0.8)}}, k\in [12,20]$ & \\
    \bottomrule

    \multicolumn{3}{l}{\scriptsize $LAI_{\text{NDVI}}$ and $LAI_{\text{IDVI}}$ are estimation results from individual statistical models for each index,} \\
    \multicolumn{3}{l}{\scriptsize applied to the appropriate LAI levels as defined by the author \citep{sun-2018}.} \\

    \end{tabular}
\end{center}
\end{table}

\subsection{Soil}
\label{sub:soil}

Numerous soil characteristics can exert a substantial influence on fire behavior and severity. However, the most relevant are the soil's organic contents~\citep{merino2021high,agbeshie2022review} and moisture levels~\citep{krueger2015soil,chaparro2016predicting}.
Higher levels of organic matter increase fuel load, making the soil more susceptible to combustion and intensifying the effects of fires. Areas with elevated organic matter content may experience deep-seated and long-lasting underground fires. On this front, a series of studies demonstrated a prediction model supported by several index combinations, including the \gls{sr}, \gls{dvi}, \gls{ndvi}, and \gls{gndvi} indexes, reporting high accuracy levels for agricultural soil mapping, with a maximum $R^2$ of 0.93 \citep{guo-2020,guo-2021}.

Dry soils contribute to faster fire spread and increased fire intensity. Conversely, moist soils can act as natural dampers, slowing the progression of fires. Simple models using the \gls{ndvi} and surface temperature for soil moisture estimation have been proposed \citep{sandholt-2002}. However, more recent studies show improved soil moisture estimation accuracy \citep{juntao-2023,swain-2021} by leveraging indices such as \gls{gndvi}~\citep{gitelson-1996}, \gls{savi}, \gls{msavi}~\citep{zhao-2022}, and \gls{osavi}~\citep{rendeaux-1996}. Other multispectral approaches, which incorporate infrared and thermal data along with more sophisticated analysis algorithms, have also reported good results for soil water content prediction in agricultural settings, including $R^2$s upwards of 0.7 \citep{bertalan-2022} and RMSEs as low as 2.5\% \citep{seo-2021}.

\subsection{Water}
\label{sub:water}

The presence of water features has two significant impacts on wildfire management: they provide natural firebreaks and serve as essential water sources for firefighting operations. Rapid identification and access to substantial volumes of water, such as lakes, rivers, and reservoirs, are essential for effective fire suppression.

Similarly to the vegetation indexes, water indexes such as the \gls{ndwi} and \gls{mndwi} \citep{zhao-2022} have their own advantages and disadvantages according to the scenario in question \citep{fangfang-2011}. A study that investigated the performance of solo implementations of the \gls{ndwi} and \gls{mndwi} reported an overall accuracy of 77\% and 84.3\%, respectively \citep{masocha-2018} in delineating land surface water features. Moreover, the same study also suggests that fusing vegetation indices with the aforementioned water indices can lead to a reduction in their overall accuracy.

Several machine learning approaches have been developed in an effort to achieve better accuracy and performance in water feature segmentation, including the WatNet \citep{luo-2021} and MC-WBDN \citep{yuan-2021} models,  as well as the \gls{muwi}\footnote{Formula available in Appendix~\ref{sub:annex-formulas}, Supplementary Table~\ref{tab:sup-indices}.}\citep{wang-2018} index. The WatNet model achieved over 95\% accuracy in water delineation, surpassing the \gls{mndwi} in three selected test regions, which included an urban landscape, a mountainous area, and a vegetated region during high cloud cover. Similarly, the MC-WBDN solution outpaced traditional indices, including \gls{ndwi}, \gls{ndmi}, and \gls{mndwi}. It achieved the highest performance among all tested methods, with a \gls{miou} of 74.42\%, indicating a high probability that a surface water pixel is correctly classified. Finally, \gls{muwi} proved its capability to generate accurate high resolution water maps. The index was validated on six testing sites, showing a higher disposition in classifying non-water as water than the contrary. It displayed overall accuracy exceeding 95\% and high statistical significance from the McNemar test, used to determine whether there is a significant difference in proportions on a binary outcome across paired or matched samples \citep{mcnemar1947note}.

\subsection{Artificial Structures}
\label{sub:artificial}

Artificial surfaces include human-made structures such as buildings, roads, bridges, and other infrastructure, which differ from natural vegetation and terrain in both composition and response to fire. This section focuses primarily on road extraction.

The \gls{rei} index has achieved accuracy levels of 86\% to 88\% in extracting asphalt road networks~\citep{shahi-2015}. While this index generally provides useful results, its performance decreases when roads are obscured by trees or shadows, and it has been observed to misclassify building boundaries as roads~\citep{shahi-2015}.

Ahmed et al. \citep{waqas-2022} have also demonstrated high-precision extraction of roads using the \gls{ri} index. Their proposal successfully detected roads with widths of less than 10 meters in certain regions and has also been able to identify gaps in road features in the presence of buildings and structures made of concrete. In turn, Zhao \& Zhu~\citep{zhao-2022} observed how the \gls{asi} index improved over seven contemporary indices\footnote{These and other less common indices are summarized in Appendix~\ref{sub:annex-formulas}, Supplementary Table~\ref{tab:sup-indices} and are not discussed in detail in this work.}--- UI \citep{kawamura-1996}, NDBI \citep{zha-2003}, IBI \citep{xu-2008}, BCI \citep{deng-2012}, VgNIR-BI \citep{estoque-2015}, PISI \citep{tian-2018} and BLFEI \citep{bouhennache-2019}---in road extraction over eight study areas, displaying a JMD, TD, and SDI between 2\% to 75\%, 13\% to 164\%, and 3\% to 131\%, respectively.

Finally, the integration of vision algorithms with single indices can enhance the extraction process and minimize miss-classifications of roads. For example, the inclusion of machine learning and a data fusion methodology based on majority voting was shown to outperform single indices such as \gls{ndvi}, \gls{ndwi}, and \gls{savi} for road extraction on 8 rural, semi-urban and urban environments, with an overall accuracy of 97.78\% \citep{puttinaovarat-2018}.

\subsection{Burnt Areas}
\label{sub:burnt}

The identification of burnt areas facilitates wildfire damage assessment---in particular the estimation of repercussions on ecosystems and infrastructures---fostering well-informed strategies for post-fire recovery, rehabilitation, and future risk mitigation. Moreover, analysis of the extent and patterns of burnt areas can provide a deeper understanding of the general behavior and characteristics of wildfires within a specific region.

The introduction of the \gls{bai} index significantly improved burnt area estimation due to its higher sensitivity compared to \gls{ndvi} and \gls{savi} \citep{chuvieco-2002}. However, due to the high internal variability of scorched terrain, \citet{chuvieco-2002} suggest using one standard deviation of the \gls{bai} as a segmentation threshold.

The introduction of the \gls{bai} index, which exhibits higher sensitivity to burnt terrain compared to \gls{ndvi} and \gls{savi}, considerably improved the estimation of burnt areas~\citep{chuvieco-2002}. However, the same study advises caution when using the index for burned land mapping due to the significant variability within scorched areas. In particular, concerning the use of thresholding techniques, the authors recommend the use of one standard deviation of \gls{bai} as a threshold, ensuring the most accurate results possible. To address landscape complexity, \citet{filipponi2018bais2} developed the \gls{bais2} using optimized Sentinel-2 \gls{msi} bands, while \citet{alcaras2022normalized} integrated traditional indices with normalized difference methods to improve mapping accuracy in specific topographies. To further mitigate interference from terrain shadows and complex surface environments, \citet{guo2024assessment} evaluated the \gls{abai}, demonstrating superior stability and accuracy in detecting forest fire severity compared to traditional metrics.

Subsequent research proposed an \gls{nbr}-based approach \citep{lopez-1991, key-2006} to detect changes in burn extent and severity using satellite imagery. Although dependent on the \gls{swir} band---which may involve higher acquisition costs---this method is highly effective for aggregating information across broad geographical and temporal scales.

Another study~\citep{key-2006} proposed an \gls{nbr}-based~\citep{lopez-1991} approach using satellite imagery to detect changes in the extent and severity of burns.  Even though it uses the \gls{swir} band---which is typically costly to acquire---the method was particularly successful in comparing results and aggregating information across wide geographical areas and over time. \citet{aksoy2023muugla} utilized the Muğla wildfires to validate multi-temporal analysis for tracking fire progression, while \citet{wu2022forest} leveraged Google Earth Engine to scale these assessments across large forest ecosystems.

\citet{holden-2005} highlighted the potential of the \gls{nbrt1} index---which directly incorporates thermal reflectance data---in identifying burned areas with low fire-induced vegetation mortality. However,  the authors emphasize the need for further research on how the timing of post-fire image acquisition affects the distinction between burned and unburned areas using two- or three-dimensional indices. To address this, \citet{liu2025tssa} introduced a \gls{tssa} to better differentiate between actual fire disturbances and seasonal phenological changes.

\subsection{Spatiotemporal and Multiscale Dynamics}
\label{sub:spatiotemporal-dynamics}

Multispectral index analysis can be divided into two approaches: \textit{static} and \textit{spatiotemporal}. Both frameworks operate under a set of multiscale constraints---the inherent trade-offs between spatial, temporal, and spectral resolutions.

Static analysis uses imagery from a single point in time to characterize land conditions a process that is largely a function of spatial grain. While high-resolution systems are capable of isolating fine-scale features, such as localized canopy gaps or individual vegetation patches, moderate-resolution sensors frequently encounter mixed pixel effects. In these instances, a single pixel represents an aggregate of distinct land cover types, leading to spectral blurring that can mask subtle landscape transitions or underestimate localized disturbances. Such scale dependency limits the transferability of threshold-based methods between different sensors without rigorous recalibration. Furthermore, while static analysis is essential for baseline mapping, it remains limited by an inability to distinguish between transient fluctuations and long-term trends; for example, a single snapshot may identify low vegetation cover but cannot determine its underlying cause or duration.

Spatiotemporal analysis extends these multiscale dynamics in its assessment of how indices fluctuate across both temporal and spatial dimensions to detect trends, seasonal patterns, and disturbance regimes, such as wildfires. This approach is primarily is contingent upon the space-time trade-off: while high spatial resolution provides necessary structural detail, it often entails lower temporal frequency and longer revisit times. Conversely, high-frequency temporal data often requires sacrificing spatial detail, causing small scale recovery patterns to be lost within coarse-grained pixels. Time series data from indices such as \gls{ndvi} or \gls{nbr} can help evaluate vegetation loss, recovery, and burn severity, yet their efficacy remains contingent upon navigating these resolution trade-offs to support accurate long-term monitoring and land management efforts.

The practical case studies discussed in the following section focus mainly on static analysis. Although spatiotemporal methods provide deeper insight, static analysis offers a faster and simpler way to assess post-fire conditions and map risks, using recent imagery and threshold-based methods without requiring complex temporal data.

\section{Practical Case Studies}
\label{sec:case-studies}

This section presents two case studies that explore the practical application of multispectral indices for terrain feature extraction. The first case study, detailed in Subsection~\ref{sub:study-1}, conducts an empirical analysis of the results obtained from directly applying selected indices to satellite imagery. The second case study, described in Subsection~\ref{sub:study-2}, extends the analysis by implementing a systematic methodology for index evaluation, facilitating the acquisition of concrete, comparable results for further interpretation.

\subsection{Study I}
\label{sub:study-1}

Study I performs a qualitative assessment of the performance of several spectral indices, including \gls{ndvi}, \gls{ndwi}, \gls{ari}, \gls{mari}, \gls{asi}, \gls{rei}, \gls{nbr}, and \gls{bai}. These indices are derived from satellite reflectance data collected across three distinct regions in Portugal. The analysis focuses on qualitatively evaluating the extent to which each index effectively highlights specific features or phenomena of interest.

\subsubsection{Data Collection}
\label{sub:data-collection-1}

Satellite imagery collection and processing were performed using the \gls{gee} platform~\citep{gorelick-2017}, facilitating the calculation of selected multispectral indices for designated regions of interest based on Landsat 8 surface reflectance data. For each site under study, a central coordinate point was selected to define the area of interest, from which a bounding square with a side length of 40 kilometers was extracted, yielding a total area of 1,600 square kilometers.

\subsubsection{Regions of Interest}
\label{sub:roi-1}

\begin{figure*}
\begin{center}
\footnotesize
\includegraphics[width=.9\textwidth]{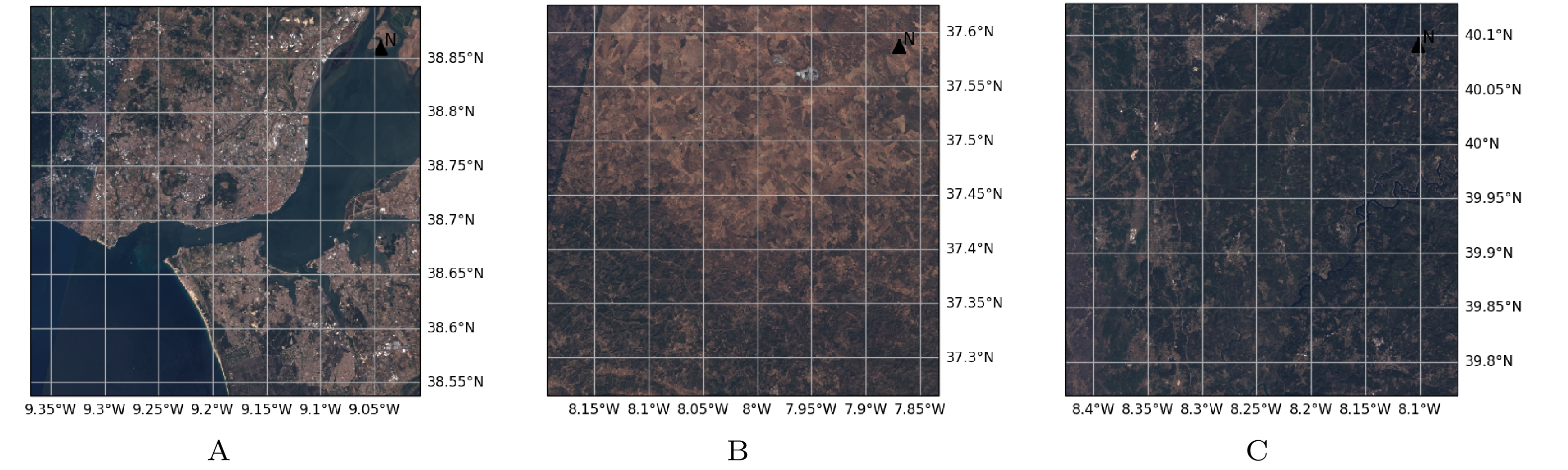} \\
\caption{RGB representation of the three regions of interest in Study I: A -- greater Lisbon area; B -- southern Portugal; and, C -- central Portugal. \label{fig:case-study-1-rgb}}
\end{center}
\end{figure*}

Three regions of interest, illustrated in Fig.~\ref{fig:case-study-1-rgb}, were selected for this study due to their diversity and abundance of relevant extractable terrain features, including vegetated areas, water bodies, and urbanized zones. Geographic details for these regions are presented in Table~\ref{tab:roi-1-geo}, and the regions themselves can be described as follows:

\begin{itemize}
  \item[A] Located within the Lisbon Metropolitan Area, this region features a complex urban and suburban landscape interspersed with significant green spaces and extensive infrastructure. Key geographical features include the Tagus Estuary on the eastern side, surrounded by urban developments, and its drainage into the Atlantic Ocean to the west. The Sintra mountain range is situated in the northwestern part of the region. Additionally, the area is traversed by two major bridges spanning the Tagus River, underscoring its infrastructural significance.

  \item[B] This region in southern Portugal represents a transitional landscape where the sparsely vegetated southern Alentejo region converges with the mountainous ranges of southern Portugal. The northern portion of the area is characterized by a relatively barren, patchwork terrain, reflective of agricultural activities interspersed with natural sparse vegetation. Progressing southward, the terrain transitions into a more rugged and mountainous landscape, with darker hues indicative of denser vegetation and forested areas. This contrast underscores the gradual shift from predominantly agricultural lands to more densely vegetated mountainous regions.

  \item[C] Located in central Portugal, this region is characterized by a relatively homogeneous distribution of vegetation. The landscape predominantly consists of a mix of dense forested patches interspersed with open areas, including agricultural lands. A river traverses the region, contributing to its geographical diversity.
\end{itemize}

\begin{table}
    \centering
    \caption{Properties of the regions of interest for Study I.}
    \label{tab:roi-1-geo}
    \begin{tabular}{llll}
        \toprule
        \multirow{2}{*}{Property} & \multicolumn{3}{c}{Regions} \\
        \cmidrule(rl){2-4}
        & A & B & C \\
        \midrule
        Latitude$^*$ & 38.7180 & 37.4453 & 39.9494 \\
        Longitude$^*$ & -9.1886 & -8.0131 & -8.2455 \\
        Area (Km$^2$) & 1600 & 1600 & 1600 \\
        \bottomrule
    \multicolumn{4}{l}{\scriptsize $^*$Coordinates are in the WGS84 standard.}\\
    \end{tabular}
\end{table}

\subsubsection{Results and Analysis}
\label{sub:results-1}

\begin{figure}
\begin{center}
\footnotesize
\includegraphics[width=.95\textwidth]{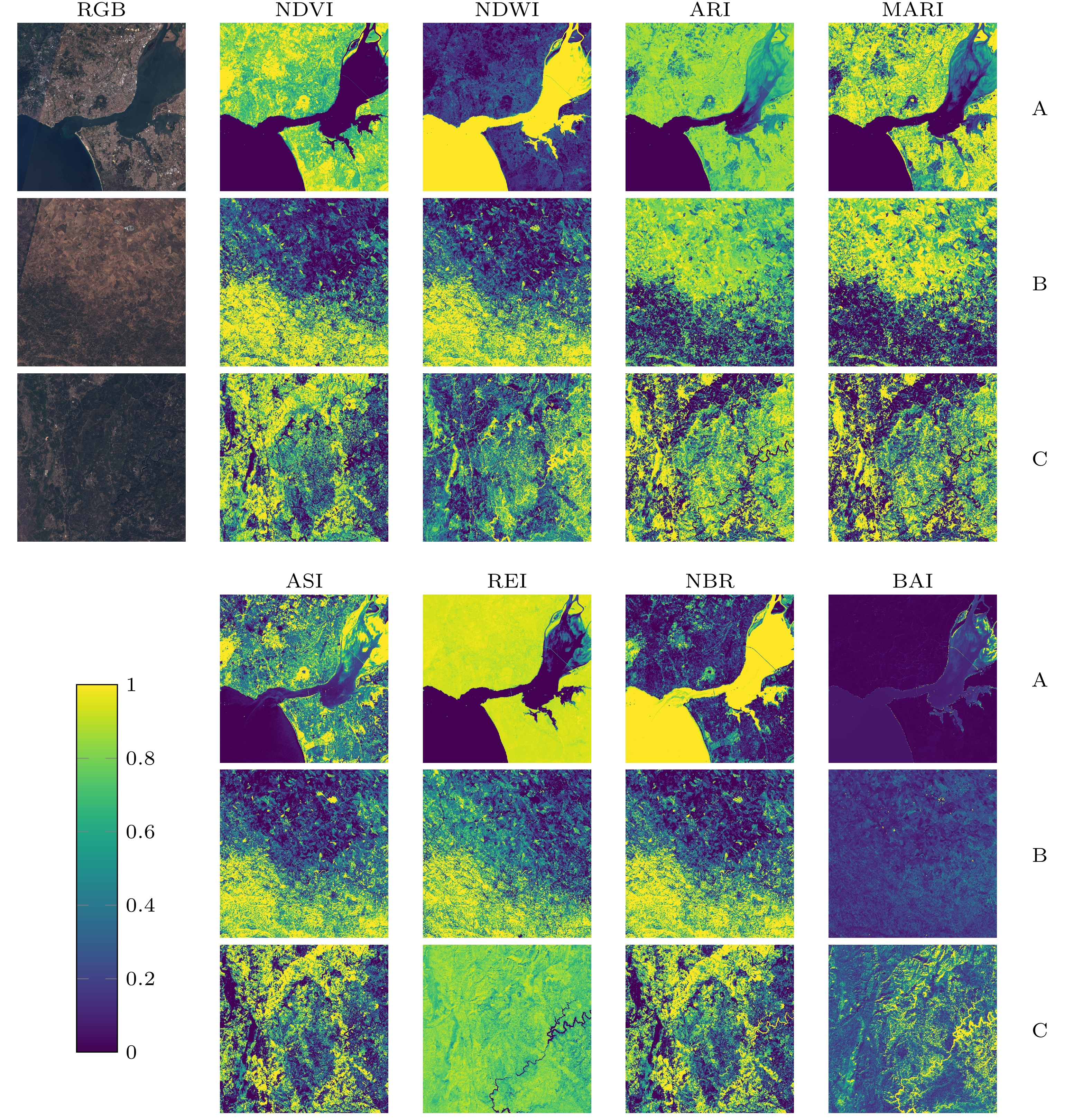}
\caption{Study I results. Multispectral indices across three test sites (A, B, and C). The upper panel displays true-color (RGB) imagery alongside NDVI, NDWI, ARI, and MARI. The lower panel displays ASI, REI, NBR, and BAI. All indices are visualized using the \textit{Viridis} colormap, with values normalized according to the provided scale bar.\label{tab:case-study-1}}
\end{center}
\end{figure}

The multispectral index maps generated for the regions of interest in Study I are presented in Figure~\ref{tab:case-study-1}, highlighting the most prevalent indices in the literature, which will be discussed below. Additional results, organized according to type, are provided in Appendix~\ref{sub:annex-figures} (Supplementary Figures~\ref{tab:practical-simple-1}–\ref{tab:practical-burnt}) for conciseness and readability.

\gls{ndvi} serves as an indicator of vegetation greenness, with higher values typically denoting dense, green vegetation, and lower values indicating sparsely vegetated or barren areas. This pattern is evident in the analyzed regions. In Region A, urban areas display low \gls{ndvi} values, reflecting limited vegetation, while green spaces such as the Sintra Mountains and Monsanto Park exhibit high \gls{ndvi} values, indicative of substantial vegetation cover. Region B demonstrates a pronounced contrast in \gls{ndvi} values, with the sparsely vegetated northern plains showing low values, and the densely vegetated southern mountainous areas showing high values. This distribution aligns with the spatial arrangement of agricultural activities and natural vegetation. In Region C, relatively homogeneous \gls{ndvi} values are observed, consistent with a landscape characterized by a mix of forested patches and agricultural lands.

\gls{ndwi} is most effectively employed to delineate water bodies and evaluate moisture content in vegetation. High \gls{ndwi} values indicate the presence of water, whereas low values correspond to dry conditions or urbanized areas. In Region A, the high \gls{ndwi} values distinctly outline the Tagus Estuary, in contrast to the low values observed in the surrounding areas, including shallow coastal zones where the riverbed's bottom soil is visible. For Region B, the \gls{ndwi} map does not reveal significant water presence in the northern areas, consistent with the prevailing arid conditions. However, it highlights the southern region, characterized by dense green vegetation. In Region C, the \gls{ndwi} map captures some green forested patches but does not prominently distinguish any major water features, excluding the river, albeit very roughly. This limitation may be attributed to insufficient reflectance differentiation between the forested areas and the water.

The \gls{ari} and \gls{mari} are used to detect the presence of anthocyanins, which impart blue, red, or purple hues to vegetation. Higher values of these indices typically indicate a significant presence of anthocyanins. However, in Regions A, B, and C, a recurring challenge arises: non-vegetative features exhibiting similar spectral characteristics to those associated with anthocyanins are more prevalent than vegetation itself. In Region A, the urban landscape predominates; in Region B, red-toned arid soils to the north are dominant; and in Region C, sparsely vegetated patches exhibit tones resembling those of anthocyanins. While the arid and sparsely vegetated areas in Regions B and C may indeed harbor vegetation rich in anthocyanins, the differentiation power of the indices at this scale proves insufficient for accurate detection.

The \gls{asi} and \gls{rei} indices are designed to focus on artificial structures, with \gls{asi} providing a general indication for urban areas and \gls{rei} targeting roads and similar infrastructure. Higher values of these indices signify the presence of artificial structures, while lower values suggest natural or vegetated areas. In Region A, the \gls{asi} index exhibits some similarity to \gls{ari} and \gls{mari}, but with enhanced differentiation within urban areas, particularly along the major thoroughfares near the coast on both sides of the estuary. Additionally, both bridges are more distinctly recognizable. The \gls{rei} index, while still emphasizing the two bridges, shows limited differentiation in the remaining urban zones. In Region B, both indices display a similar north-south intensity gradient; however, the differentiation between regions is more pronounced with the \gls{asi}. Notably, no significant urban areas are distinguished in this region. In Region C, \gls{asi} highlights the major vegetated areas, but fails to distinguish any significant urban zones or infrastructure. \gls{rei} presents relatively uniform values across most of the region, except for the bottom-right quadrant, where it clearly delineates the area along the river. \gls{asi} shows strong differentiation capabilities for urban areas, particularly large thoroughfares, but its sensitivity to vegetation in proximity to urban zones limits its ability to effectively distinguish these areas. Conversely, \gls{rei} seems to be overshadowed by the presence of water bodies, such as the Tagus Estuary in Region A and the river in Region C, which diminishes its effectiveness in delineating roads. In Region B, where no significant water bodies are present, \gls{rei} results align more closely with those of other similar indices.

The \gls{nbr} and \gls{bai} indices are employed to assess fire severity by delineating burned areas. Generally, low values correspond to minimal recent burn activity, whereas higher values indicate the presence of recent fires. Although no recent fire activity is observed in Regions A, B, and C, these indices can still provide valuable insights based on the features they highlight. Burnt areas typically appear dark in multispectral imagery, and these indices are designed to capture such tonalities effectively. The \gls{nbr} index successfully outlines the darkest features of the imagery across all three regions. In Region A, it emphasizes the water bodies and dense vegetation; in Region B, it accentuates the vegetated southern region; and in Region C, it highlights the dense patches of vegetation and the river. Conversely, the \gls{bai} index produces a more homogeneous output. In Region A, it highlights bottom soil near the river shallows; in Region B, it emphasizes the vegetation across the southern region; and in Region C, it functions similarly to \gls{nbr}, although it seems to capture the darkest areas of the map with greater precision.

\subsection{Study II}
\label{sub:study-2}

Study II expands upon the findings of the first study by quantitatively evaluating the effectiveness of various multispectral indices in extracting vegetation, water bodies, and artificial structures from satellite imagery. For each region of interest, a terrain feature map was generated to serve as the ground truth, providing a reliable benchmark for comparison with the processed data. This methodology facilitates the extraction of concrete and statistically significant results, enabling a more objective and robust analysis.

The analysis of multispectral index maps entails calculating the distribution of values within predefined thresholds for a specific extractable feature, using terrain feature maps as masks. This optimal range was identified heuristically based on the value distribution of feature-coincident pixels, employing a Gaussian filter for smoothing, as highlighted in Equation~\ref{eq:gaussmoth}:

\begin{equation}
\label{eq:gaussmoth}
    R^* = \arg\max_{l \leq m \leq r} \left\{ \sum_{i=l}^{r} \tilde{D}_i \;\middle|\; \sum_{i=l}^{r} \tilde{D}_i > (1 + \epsilon) \cdot \tilde{D}_m \right\}
\end{equation}

\noindent where \( D \) denotes the original distribution of pixel values, \( \tilde{D} \) is the smoothed distribution obtained via a Gaussian filter with standard deviation \( \sigma = 1.0 \), \( m \) represents the index of the maximum value in \( \tilde{D} \), \( \epsilon \) denotes the minimum required relative increase (5\%) in the cumulative sum of the distribution to justify expanding the range, and \( R^* = (t_l, t_r) \) corresponds to the selected optimal threshold range in the domain of the original value axis \( T = \{t_1, \ldots, t_n\} \).

\subsubsection{Data Collection}
\label{sub:data-collection-2}

Two resources were obtained from the Portuguese Territory Authority\footnote{Direção Geral do Território: \url{https://www.dgterritorio.gov.pt/}.}: the official district geographic shapefiles and the 2022 edition of the official Portuguese land use map \citep{SNIG2024}. These resources enabled the extraction of land use maps for each district. Additionally, satellite reflectance data from the Sentinel-2\footnote{ESA Sentinel-2: \url{https://www.esa.int/Applications/Observing_the_Earth/Copernicus/Sentinel-2}.} platform were retrieved through the OpenEO API \citep{lahn2025}, providing the necessary data for computing the multispectral indices for each district.

\subsubsection{Regions of Interest}
\label{sub:roi-2}

\begin{figure*}
\begin{center}
\footnotesize
\includegraphics[width=.85\textwidth]{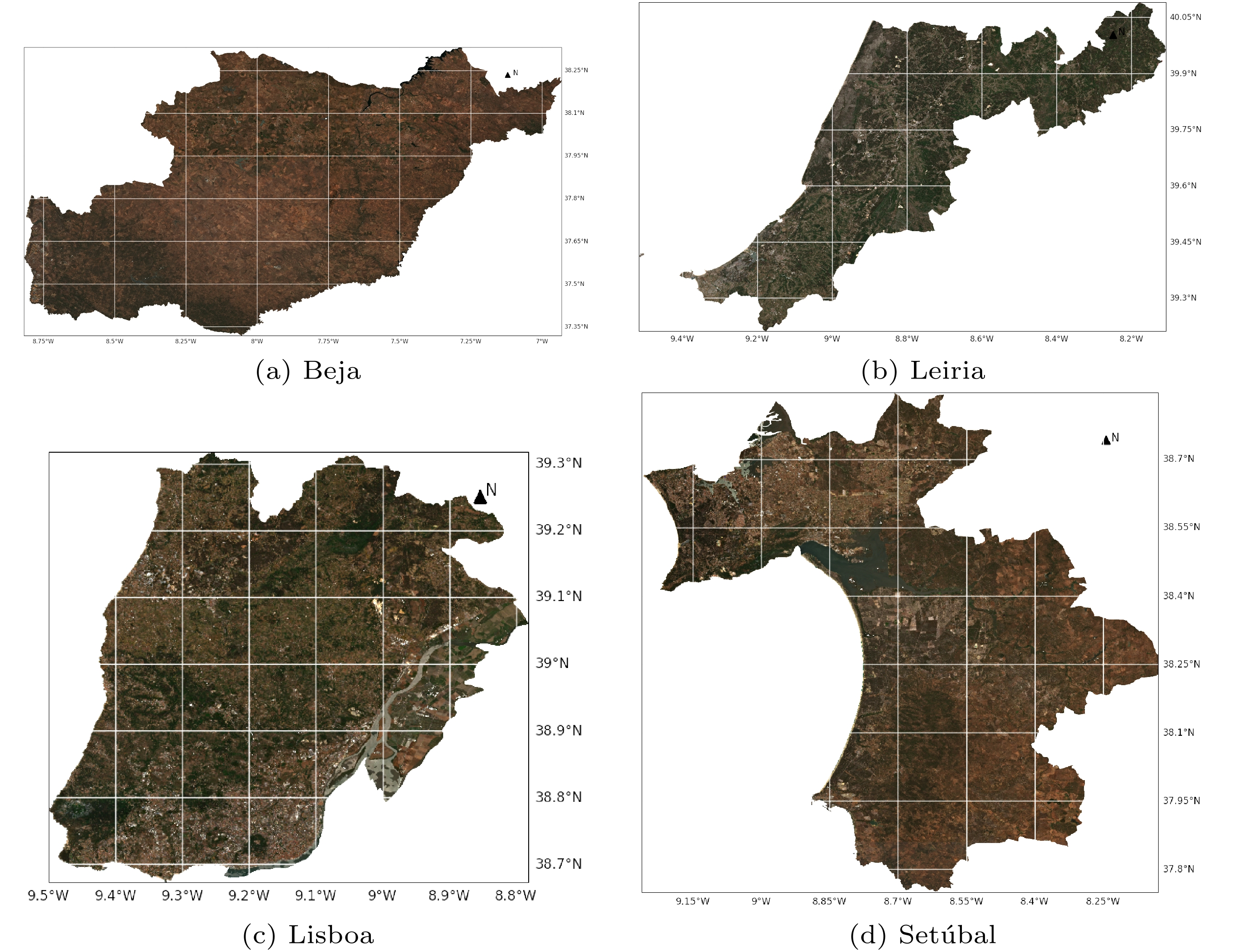}
\caption{RGB representation of the four regions of interest: the districts of (a) Beja, (b) Leiria, (c) Lisboa and (d) Setúbal. \label{fig:case-study-2-rgb}}
\end{center}
\end{figure*}

Portugal is composed of 18 districts  (and 2 autonomous regions); from these, four districts, depicted in Figure~\ref{fig:case-study-2-rgb}, were selected based on similar criteria and their approximate geographic proximity to the regions of interest in Study I. These districts, detailed in Table~\ref{tab:roi-2-geo} and depicted (RGB) in Figure~\ref{fig:case-study-2-rgb}, can be described as follows:

\begin{itemize}
    \item \textbf{Beja} (Figure~\ref{fig:case-study-2-rgb}a): Partially depicted in the northern half of Region B, the Beja district is characterized by vast semi-arid plains, with notable expanses dedicated to viticulture and agriculture.
    \item \textbf{Leiria}  (Figure~\ref{fig:case-study-2-rgb}b): Encompassing a portion of Region C, the Leiria district presents a diverse landscape, featuring a coastline, extensive forested areas, and expansive agricultural plains.
    \item \textbf{Lisbon and Setúbal}  (Figures~\ref{fig:case-study-2-rgb}c and ~\ref{fig:case-study-2-rgb}d): These two districts, partially depicted in Region A, are characterized by a blend of urban and suburban areas. Furthermore, they are distinguished by significant coastal regions, natural reserves, and extensive forested areas.
\end{itemize}

\begin{table}
    \caption{Properties of the regions of interest for Study II.}
    \label{tab:roi-2-geo}
    \centering
    \begin{tabular}{lrrrr}
        \toprule
        Property & A & B & C & D \\
        \midrule
        Name & Beja & Leiria & Lisboa & Setúbal \\
        Latitude$^a$ &  38.0167 & 39.75 & 38.7253 & 38.5245 \\
        Longitude$^a$ & -7.8667 & -8.8 & -9.15 & -8.8931 \\
        Area (Km$^2$) & 10263 & 3515 & 2761 & 5064 \\
        Population$^b$ & 147 & 467 & 2301 & 887 \\
        \midrule
        Vegetation$^c$ & 51.06\% & 26.41\% & 50.89\% & 36.75\% \\
        Water$^c$ & 0.72\% & 0.13\% & 2.07\% & 2.40\% \\
        Artificial Structures$^c$ & 0.30\% & 1.5\% & 7.74\% & 1.64\% \\
        \bottomrule
        \multicolumn{4}{l}{\scriptsize $^a$Coordinates are in the WGS84 standard.}\\
        \multicolumn{4}{l}{\scriptsize $^b$Hundreds of thousands, as of 2022.}\\
        \multicolumn{4}{l}{\scriptsize $^c$Area as percentage of entire region.}\\
    \end{tabular}
\end{table}

\subsubsection{Results and Analysis}
\label{sub:results-2}

The accuracy results for the distribution of vegetation, water, and artificial structures are presented in Tables~\ref{tab:acc-indices-v},~\ref{tab:acc-indices-w}, and~\ref{tab:acc-indices-as}, respectively. The corresponding segmentation threshold values are provided in Table~\ref{tab:ran-indices}.

\begin{table}
\begin{center}
\caption{Accuracy percentage for optimal threshold ranges in extracting vegetation distribution for each index and region of interest.\label{tab:acc-indices-v}}
\small
\begin{tabular}{lrrrrrr}
    \toprule
    \multirow{2}{*}{Index} & \multicolumn{4}{c}{Regions} & \multicolumn{2}{c}{Statistics} \\
    \cmidrule(rl){2-5} \cmidrule(rl){6-7}
    & Beja & Leiria & Lisboa & Setúbal & $\bar{X}$ & $\sigma$ \\
    \midrule
    SR & 56.08 & 11.00 & 38.10 & 46.63 & 37.95 & 16.81 \\
    NDVI & 95.85 & 89.06 & 78.36 & 85.23 & 87.12 & 6.33 \\
    DVI & 54.14 & 12.80 & 64.31 & 61.53 & 48.20 & 20.77 \\
    RDVI & 79.73 & 65.94 & 78.18 & 74.38 & 74.56 & 5.34 \\
    MSR & 45.11 & 10.55 & 47.81 & 47.42 & 37.72 & 15.72 \\
    GNDVI & 94.70 & 89.16 & 84.56 & 90.09 & 89.63 & 3.60 \\
    GARI & 95.52 & 88.17 & 82.51 & 85.01 & 87.80 & 4.89 \\
    NDRE & 95.07 & 90.91 & 81.22 & 88.70 & 88.98 & 5.03 \\
    GDVI & 52.53 & 45.69 & 61.96 & 61.75 & 55.48 & 6.82 \\
    GRVI & 59.17 & 13.43 & 45.40 & 60.39 & 44.60 & 18.93 \\
    IDVI & 93.18 & 79.50 & 82.88 & 89.95 & 86.38 & 5.44 \\
    ARI & 72.23 & 66.79 & 66.39 & 72.39 & 69.45 & 2.86 \\
    MARI & 43.14 & 10.89 & 53.08 & 52.39 & 39.88 & 17.19 \\
    EVI & 87.92 & 79.76 & 81.71 & 85.75 & 83.78 & 3.22 \\
    EVI2 & 87.96 & 82.00 & 67.44 & 76.51 & 78.48 & 7.55 \\
    SAVI & 93.17 & 88.64 & 84.76 & 84.19 & 87.69 & 3.60 \\
    MSAVI & 52.79 & 54.05 & 53.86 & 56.76 & 54.36 & 1.46 \\
    OSAVI & 95.44 & 89.62 & 85.52 & 86.20 & 89.20 & 3.93 \\
    MBI & 78.48 & 57.18 & 63.33 & 74.04 & 68.26 & 8.44 \\
    EMBI & 67.59 & 64.92 & 69.62 & 65.28 & 66.85 & 1.90 \\
    TAVI & 55.49 & 10.79 & 43.50 & 46.51 & 39.07 & 16.91 \\
    NDWI & 60.59 & 10.60 & 57.72 & 60.11 & 47.26 & 21.19 \\
    NDMI & 89.57 & 79.09 & 63.75 & 35.52 & 66.98 & 20.35 \\
    MNDWI & 53.55 & 10.56 & 54.88 & 14.34 & 33.33 & 20.93 \\
    NBR & 89.57 & 79.09 & 63.75 & 35.52 & 66.98 & 20.35 \\
    BAI & 95.36 & 89.63 & 81.00 & 85.85 & 87.96 & 5.25 \\
    AF & 95.86 & 87.92 & 86.52 & 91.04 & 90.34 & 3.59 \\
    VSF & 52.87 & 11.05 & 47.18 & 61.30 & 43.10 & 19.17 \\
    SSF & 84.44 & 77.03 & 77.20 & 78.19 & 79.22 & 3.05 \\
    MF & 74.05 & 61.14 & 63.67 & 68.96 & 66.96 & 4.97 \\
    ASI & 95.35 & 89.69 & 77.73 & 85.74 & 87.13 & 6.41 \\
    REI & 70.00 & 61.88 & 62.51 & 69.42 & 65.95 & 3.77 \\
    RI & 95.31 & 89.58 & 81.00 & 85.85 & 87.94 & 5.23 \\
    \bottomrule

    \end{tabular}
\end{center}
\end{table}

\begin{table}
\begin{center}
\caption{Accuracy percentage for optimal threshold ranges in extracting water distribution for each index and region of interest.\label{tab:acc-indices-w}}
\small
\begin{tabular}{lrrrrrr}
    \toprule
    \multirow{2}{*}{Index} & \multicolumn{4}{c}{Regions} & \multicolumn{2}{c}{Statistics} \\
    \cmidrule(rl){2-5} \cmidrule(rl){6-7}
    & Beja & Leiria & Lisboa & Setúbal & $\bar{X}$ & $\sigma$ \\
    \midrule
    SR & 99.02 & 99.36 & 99.13 & 98.40 & 98.98 & 0.36 \\
    NDVI & 98.65 & 99.55 & 96.71 & 94.44 & 97.34 & 1.96 \\
    DVI & 98.64 & 98.97 & 96.71 & 94.40 & 97.18 & 1.82 \\
    RDVI & 98.65 & 99.55 & 96.70 & 94.40 & 97.32 & 1.98 \\
    MSR & 96.35 & 99.17 & 63.59 & 67.60 & 81.68 & 16.18 \\
    GNDVI & 98.66 & 99.55 & 96.71 & 94.42 & 97.34 & 1.97 \\
    GARI & 98.66 & 99.55 & 96.70 & 94.44 & 97.34 & 1.97 \\
    NDRE & 98.66 & 99.54 & 96.69 & 94.42 & 97.33 & 1.97 \\
    GDVI & 98.65 & 99.33 & 96.71 & 94.41 & 97.28 & 1.91 \\
    GRVI & 99.31 & 99.68 & 99.03 & 98.19 & 99.05 & 0.55 \\
    IDVI & 98.65 & 99.55 & 96.71 & 94.41 & 97.33 & 1.97 \\
    ARI & 98.81 & 99.51 & 98.12 & 97.55 & 98.50 & 0.74 \\
    MARI & 98.93 & 97.63 & 88.11 & 92.23 & 94.22 & 4.33 \\
    EVI & 98.65 & 99.55 & 96.73 & 94.42 & 97.34 & 1.97 \\
    EVI2 & 98.66 & 99.55 & 96.71 & 94.44 & 97.34 & 1.96 \\
    SAVI & 98.66 & 99.55 & 96.71 & 94.43 & 97.34 & 1.97 \\
    MSAVI & 99.34 & 99.70 & 98.63 & 97.58 & 98.81 & 0.81 \\
    OSAVI & 98.66 & 99.55 & 96.75 & 94.48 & 97.36 & 1.95 \\
    MBI & 19.75 & 99.39 & 96.70 & 94.49 & 77.58 & 33.43 \\
    EMBI & 98.63 & 99.52 & 96.68 & 94.35 & 97.29 & 1.99 \\
    TAVI & 98.64 & 99.16 & 60.63 & 94.36 & 88.20 & 16.02 \\
    NDWI & 91.94 & 99.70 & 99.18 & 98.45 & 97.32 & 3.14 \\
    NDMI & 98.65 & 99.53 & 27.96 & 66.36 & 73.12 & 29.30 \\
    MNDWI & 98.50 & 99.39 & 98.54 & 98.12 & 98.64 & 0.46 \\
    NBR & 98.65 & 99.53 & 27.96 & 66.36 & 73.12 & 29.30 \\
    BAI & 1.30 & 0.40 & 3.29 & 5.59 & 2.64 & 2.00 \\
    AF & 98.66 & 99.55 & 96.71 & 94.45 & 97.34 & 1.96 \\
    VSF & 98.65 & 99.14 & 94.04 & 94.49 & 96.58 & 2.33 \\
    SSF & 12.76 & 99.68 & 98.80 & 95.95 & 76.80 & 37.00 \\
    MF & 28.51 & 81.87 & 98.26 & 67.14 & 68.95 & 25.81 \\
    ASI & 1.13 & 0.69 & 3.39 & 6.11 & 2.83 & 2.15 \\
    REI & 68.45 & 99.54 & 89.85 & 94.40 & 88.06 & 11.83 \\
    RI & 1.35 & 0.45 & 3.30 & 5.61 & 2.68 & 1.98 \\
    \bottomrule

    \end{tabular}
\end{center}
\end{table}

\begin{table}
\begin{center}
\caption{Accuracy percentage for optimal threshold ranges in extracting artificial structure distribution for each index and region of interest.\label{tab:acc-indices-as}}
\small
\begin{tabular}{lrrrrrr}
    \toprule
    \multirow{2}{*}{Index} & \multicolumn{4}{c}{Regions} & \multicolumn{2}{c}{Statistics} \\
    \cmidrule(rl){2-5} \cmidrule(rl){6-7}
    & Beja & Leiria & Lisboa & Setúbal & $\bar{X}$ & $\sigma$ \\
    \midrule
    SR & 66.08 & 88.18 & 80.58 & 79.59 & 78.61 & 7.96 \\
    NDVI & 8.16 & 76.46 & 70.06 & 56.00 & 52.67 & 26.74 \\
    DVI & 52.15 & 74.22 & 70.64 & 55.34 & 63.09 & 9.50 \\
    RDVI & 33.80 & 83.97 & 73.69 & 65.48 & 64.24 & 18.75 \\
    MSR & 95.39 & 94.70 & 87.72 & 95.98 & 93.45 & 3.34 \\
    GNDVI & 11.13 & 55.82 & 65.78 & 55.13 & 46.96 & 21.11 \\
    GARI & 13.53 & 72.95 & 69.34 & 50.30 & 51.53 & 23.57 \\
    NDRE & 2.69 & 71.12 & 61.18 & 28.75 & 40.94 & 27.07 \\
    GDVI & 42.44 & 54.98 & 57.51 & 55.81 & 52.68 & 5.98 \\
    GRVI & 67.17 & 85.71 & 82.71 & 84.13 & 79.93 & 7.44 \\
    IDVI & 4.08 & 49.01 & 20.78 & 8.49 & 20.59 & 17.51 \\
    ARI & 51.53 & 71.85 & 70.80 & 66.33 & 65.13 & 8.12 \\
    MARI & 57.85 & 94.64 & 87.60 & 90.60 & 82.67 & 14.55 \\
    EVI & 19.75 & 70.66 & 23.50 & 14.85 & 32.19 & 22.42 \\
    EVI2 & 27.88 & 87.17 & 77.70 & 63.23 & 64.00 & 22.53 \\
    SAVI & 15.94 & 79.39 & 70.19 & 56.53 & 55.51 & 24.25 \\
    MSAVI & 44.41 & 55.27 & 46.17 & 47.87 & 48.43 & 4.13 \\
    OSAVI & 9.95 & 77.02 & 66.16 & 41.68 & 48.70 & 25.78 \\
    MBI & 25.68 & 65.57 & 47.37 & 32.69 & 42.83 & 15.29 \\
    EMBI & 45.34 & 65.33 & 44.26 & 40.97 & 48.98 & 9.58 \\
    TAVI & 60.50 & 88.89 & 79.99 & 75.53 & 76.23 & 10.28 \\
    NDWI & 56.01 & 53.98 & 52.23 & 56.50 & 54.68 & 1.70 \\
    NDMI & 37.13 & 34.84 & 33.41 & 64.53 & 42.48 & 12.80 \\
    MNDWI & 55.32 & 72.79 & 62.53 & 59.82 & 62.62 & 6.41 \\
    NBR & 37.13 & 34.84 & 33.41 & 64.53 & 42.48 & 12.80 \\
    BAI & 0.62 & 5.12 & 12.32 & 3.82 & 5.47 & 4.28 \\
    AF & 7.28 & 43.35 & 54.80 & 40.30 & 36.43 & 17.68 \\
    VSF & 58.84 & 84.41 & 74.69 & 71.31 & 72.31 & 9.14 \\
    SSF & 20.87 & 42.15 & 34.19 & 23.09 & 30.08 & 8.61 \\
    MF & 28.53 & 48.42 & 39.49 & 28.97 & 36.35 & 8.23 \\
    ASI & 1.46 & 5.36 & 42.36 & 4.54 & 13.43 & 16.77 \\
    REI & 28.36 & 44.46 & 41.89 & 38.14 & 38.21 & 6.12 \\
    RI & 0.57 & 5.08 & 12.32 & 3.83 & 5.45 & 4.29 \\
    \bottomrule

    \end{tabular}
\end{center}
\end{table}

\clearpage
\begin{landscape}
\begin{table}
\begin{center}
\caption{Optimal threshold ranges in extracting artificial structures, vegetation, or water data for each index and region of interest.\label{tab:ran-indices}}
\resizebox{\linewidth}{!}{
\begin{tabular}{lcccccccccccc}
    \toprule
    \multirow{2}{*}{Index} & \multicolumn{4}{c}{Vegetation} & \multicolumn{4}{c}{Water} & \multicolumn{4}{c}{Artificial Structures} \\
    \cmidrule(rl){2-5} \cmidrule(rl){6-9} \cmidrule(rl){10-13}
    & Beja & Leiria & Lisboa & Setúbal & Beja & Leiria & Lisboa & Setúbal & Beja & Leiria & Lisboa & Setúbal \\
    \midrule
    SR & [-0.01,0.15] & [0.98,1.0] & [-0.16,0.06] & [0.04,0.22] & [-0.26,-0.22] & [-0.18,-0.04] & [-0.23,-0.17] & [-0.16,-0.04] & [-0.03,0.09] & [-0.06,0.06] & [-0.16,-0.04] & [-0.02,0.12] \\
    NDVI & [-0.03,0.01] & [0.02,0.14] & [-0.1,0.06] & [-0.05,0.03] & [0.98,1.0] & [0.98,1.0] & [0.98,1.0] & [0.98,1.0] & [-0.04,0.0] & [-0.04,0.04] & [-0.12,-0.02] & [-0.07,-0.01] \\
    DVI & [0.0,0.22] & [0.9,1.0] & [-0.07,0.21] & [-0.04,0.2] & [0.98,1.0] & [0.98,1.0] & [0.98,1.0] & [0.98,1.0] & [-0.17,0.11] & [-0.05,0.27] & [-0.2,0.0] & [-0.11,0.11] \\
    RDVI & [-0.06,0.06] & [0.16,0.46] & [-0.07,0.15] & [-0.09,0.09] & [0.98,1.0] & [0.98,1.0] & [0.98,1.0] & [0.98,1.0] & [-0.07,0.03] & [-0.08,0.14] & [-0.16,-0.02] & [-0.11,0.01] \\
    MSR & [0.13,0.49] & [0.98,1.0] & [-0.29,-0.05] & [-0.1,0.22] & [-0.34,-0.18] & [0.98,1.0] & [-0.02,0.3] & [-0.05,0.19] & [0.72,1.0] & [0.98,1.0] & [0.98,1.0] & [0.98,1.0] \\
    GNDVI & [-0.03,0.03] & [0.04,0.14] & [-0.05,0.07] & [-0.03,0.05] & [0.98,1.0] & [0.98,1.0] & [0.98,1.0] & [0.98,1.0] & [-0.04,0.02] & [-0.02,0.08] & [-0.1,0.0] & [-0.07,0.01] \\
    GARI & [-0.04,0.02] & [0.01,0.11] & [-0.08,0.06] & [-0.07,0.03] & [0.98,1.0] & [0.98,1.0] & [0.98,1.0] & [0.98,1.0] & [-0.04,0.0] & [-0.05,0.03] & [-0.11,-0.03] & [-0.09,-0.01] \\
    NDRE & [-0.03,0.01] & [0.0,0.1] & [-0.08,0.02] & [-0.05,0.01] & [0.98,1.0] & [0.98,1.0] & [0.98,1.0] & [0.98,1.0] & [-0.03,0.01] & [-0.03,0.03] & [-0.09,-0.03] & [-0.07,-0.01] \\
    GDVI & [0.12,0.32] & [0.17,0.49] & [-0.04,0.22] & [0.04,0.26] & [0.98,1.0] & [0.98,1.0] & [0.98,1.0] & [0.98,1.0] & [0.0,0.28] & [0.17,0.49] & [-0.16,0.1] & [-0.15,0.15] \\
    GRVI & [0.07,0.29] & [0.9,1.0] & [-0.08,0.18] & [0.14,0.42] & [-0.41,-0.23] & [-0.21,-0.09] & [-0.31,-0.23] & [-0.21,-0.11] & [-0.04,0.18] & [-0.01,0.19] & [-0.16,0.02] & [-0.02,0.18] \\
    IDVI & [-0.08,-0.04] & [-0.18,-0.1] & [-0.08,-0.04] & [-0.09,-0.05] & [0.98,1.0] & [0.98,1.0] & [0.98,1.0] & [0.98,1.0] & [-0.08,-0.04] & [-0.19,-0.13] & [-0.09,-0.05] & [-0.09,-0.05] \\
    ARI & [0.22,0.42] & [0.41,0.65] & [0.17,0.39] & [0.3,0.54] & [-0.37,-0.33] & [-0.23,-0.19] & [-0.23,-0.03] & [-0.09,0.07] & [0.07,0.31] & [0.2,0.44] & [-0.05,0.19] & [0.13,0.35] \\
    MARI & [0.23,0.55] & [0.95,1.0] & [-0.18,0.12] & [0.1,0.46] & [-0.34,-0.24] & [-0.31,-0.05] & [-0.34,-0.12] & [-0.23,-0.07] & [0.09,0.41] & [0.98,1.0] & [0.82,1.0] & [0.73,1.0] \\
    EVI & [-0.03,0.07] & [0.0,0.1] & [-0.02,0.04] & [-0.02,0.04] & [-1.0,-0.98] & [-1.0,-0.98] & [-1.0,-0.98] & [-1.0,-0.98] & [-0.05,0.03] & [-0.12,0.04] & [-0.04,0.02] & [-0.03,0.03] \\
    EVI2 & [-0.04,0.02] & [0.05,0.29] & [-0.1,0.08] & [-0.09,0.05] & [0.98,1.0] & [0.98,1.0] & [0.98,1.0] & [0.98,1.0] & [-0.04,0.0] & [-0.08,0.04] & [-0.14,-0.04] & [-0.09,-0.01] \\
    SAVI & [-0.03,0.01] & [0.03,0.17] & [-0.07,0.07] & [-0.05,0.03] & [0.98,1.0] & [0.98,1.0] & [0.98,1.0] & [0.98,1.0] & [-0.04,0.0] & [-0.05,0.05] & [-0.1,-0.02] & [-0.07,-0.01] \\
    MSAVI & [0.27,0.55] & [0.31,0.61] & [0.13,0.39] & [0.31,0.61] & [-0.36,-0.3] & [-0.24,-0.18] & [-0.48,-0.28] & [-0.34,-0.18] & [0.25,0.55] & [0.45,0.73] & [0.11,0.41] & [0.3,0.6] \\
    OSAVI & [-0.03,0.01] & [0.02,0.14] & [-0.05,0.05] & [-0.05,0.03] & [0.98,1.0] & [0.98,1.0] & [0.98,1.0] & [0.98,1.0] & [-0.04,0.0] & [-0.04,0.04] & [-0.09,-0.01] & [-0.06,0.0] \\
    MBI & [-0.04,0.02] & [0.14,0.34] & [-0.06,0.08] & [-0.05,0.03] & [-0.12,0.02] & [0.98,1.0] & [0.98,1.0] & [0.98,1.0] & [-0.05,0.01] & [0.12,0.26] & [-0.07,0.05] & [-0.04,0.02] \\
    EMBI & [-0.3,-0.06] & [-0.61,-0.31] & [-0.33,-0.11] & [-0.33,-0.11] & [0.98,1.0] & [0.98,1.0] & [0.98,1.0] & [0.98,1.0] & [-0.26,-0.08] & [-0.43,-0.23] & [-0.25,-0.05] & [-0.32,-0.12] \\
    TAVI & [-0.09,0.07] & [0.98,1.0] & [-0.16,0.08] & [-0.08,0.12] & [0.98,1.0] & [0.98,1.0] & [-0.04,0.22] & [0.98,1.0] & [-0.12,0.02] & [-0.11,0.03] & [-0.22,-0.06] & [-0.13,0.03] \\
    NDWI & [-0.01,0.19] & [0.98,1.0] & [-0.04,0.18] & [0.03,0.23] & [-0.25,0.03] & [-0.15,0.01] & [-0.29,-0.21] & [-0.18,-0.1] & [-0.03,0.13] & [0.14,0.34] & [-0.07,0.13] & [-0.01,0.17] \\
    NDMI & [-0.06,0.1] & [-0.05,0.03] & [-0.14,-0.04] & [-0.12,-0.06] & [0.98,1.0] & [0.98,1.0] & [-0.15,-0.03] & [-0.11,-0.05] & [-0.02,0.06] & [-0.05,0.01] & [-0.14,-0.04] & [-0.12,-0.06] \\
    MNDWI & [-0.06,0.14] & [0.98,1.0] & [-0.02,0.2] & [0.98,1.0] & [0.98,1.0] & [0.98,1.0] & [-0.27,-0.21] & [-0.17,-0.07] & [-0.08,0.1] & [0.08,0.24] & [-0.04,0.12] & [-0.04,0.14] \\
    NBR & [-0.06,0.1] & [-0.05,0.03] & [-0.14,-0.04] & [-0.12,-0.06] & [0.98,1.0] & [0.98,1.0] & [-0.15,-0.03] & [-0.11,-0.05] & [-0.02,0.06] & [-0.05,0.01] & [-0.14,-0.04] & [-0.12,-0.06] \\
    BAI & [-0.33,-0.29] & [-0.54,-0.5] & [-0.28,-0.24] & [-0.41,-0.37] & [-0.33,-0.29] & [-0.54,-0.5] & [-0.28,-0.24] & [-0.41,-0.37] & [-0.33,-0.29] & [-0.54,-0.5] & [-0.28,-0.24] & [-0.41,-0.37] \\
    AF & [-0.03,0.03] & [0.04,0.12] & [-0.03,0.09] & [-0.02,0.06] & [0.98,1.0] & [0.98,1.0] & [0.98,1.0] & [0.98,1.0] & [-0.04,0.02] & [-0.02,0.08] & [-0.1,0.02] & [-0.06,0.02] \\
    VSF & [-0.04,0.22] & [0.97,1.0] & [-0.15,0.09] & [-0.06,0.18] & [-1.0,-0.98] & [0.98,1.0] & [0.39,0.59] & [0.96,1.0] & [-0.18,0.1] & [-0.15,0.09] & [-0.16,-0.02] & [-0.11,0.05] \\
    SSF & [0.27,0.37] & [0.44,0.56] & [0.23,0.33] & [0.34,0.46] & [0.28,0.38] & [-0.17,-0.03] & [-0.34,-0.24] & [-0.21,-0.05] & [0.28,0.36] & [0.45,0.53] & [0.2,0.3] & [0.34,0.46] \\
    MF & [-0.07,0.05] & [0.03,0.19] & [-0.04,0.08] & [-0.05,0.05] & [-0.08,0.04] & [-0.09,0.09] & [-0.11,-0.05] & [-0.06,0.0] & [-0.06,0.04] & [0.04,0.16] & [-0.05,0.07] & [-0.03,0.05] \\
    ASI & [-0.02,0.02] & [-0.01,0.03] & [-0.03,0.03] & [-0.02,0.02] & [-0.03,0.03] & [-0.02,0.02] & [-0.03,0.05] & [-0.02,0.02] & [-0.02,0.02] & [-0.02,0.02] & [-0.03,0.01] & [-0.02,0.02] \\
    REI & [-0.08,0.0] & [-0.06,0.04] & [-0.09,-0.01] & [-0.08,0.02] & [-0.1,-0.04] & [0.98,1.0] & [0.06,0.28] & [0.98,1.0] & [-0.08,0.0] & [-0.08,0.02] & [-0.1,-0.02] & [-0.08,0.0] \\
    RI & [0.29,0.33] & [0.5,0.54] & [0.24,0.28] & [0.37,0.41] & [0.29,0.33] & [0.5,0.54] & [0.24,0.28] & [0.37,0.41] & [0.29,0.33] & [0.5,0.54] & [0.24,0.28] & [0.37,0.41] \\
    \bottomrule

    \end{tabular}}
\end{center}
\end{table}
\end{landscape}
\clearpage

Among the simple greenness indicators (\gls{sr} to \gls{idvi} in Tables~\ref{tab:acc-indices-v}--\ref{tab:ran-indices}), \gls{ndvi}, \gls{rdvi}, \gls{gndvi}, \gls{gari}, \gls{ndre}, and \gls{idvi} yielded the best results for vegetation, demonstrating superior accuracy and stability compared to the remaining indices. For water, all greenness indices performed exceptionally well, achieving accuracy values above 97\% and standard deviations below 2\%, with the exception of \gls{msr}, which showed reduced accuracy and markedly higher variability. Conversely, for artificial structures, \gls{msr} was the most effective index, delivering high accuracy and very low variance, while the remaining indices suffered from either low accuracy, high variability, or, in most cases, both.

The \gls{ari} and its modified counterpart, \gls{mari}, are designed to accentuate anthocyanin content in vegetation, which is responsible for imparting blue, red, or purple hues. While both indices yielded consistent results overall, \gls{ari} outperformed \gls{mari}, achieving higher accuracy and exhibiting lower variability for vegetation (70$\pm$3\% compared to 40$\pm$17\%) and water (99$\pm$1\% compared to 94$\pm$4\%). Particularly for vegetation, it is expected that these indices would show weaker results at this scale, especially during the summer months, when anthocyanin contents are low. For artificial structures, \gls{mari} demonstrated better accuracy; however, it was accompanied by significantly higher variability (65$\pm$8\% compared to 83$\pm$15\%).

The \gls{evi} and its two-band counterpart, \gls{evi2}, both optimized for vegetation analysis, exhibited their best performance for water ($\approx$95$\pm$2\%), closely followed by vegetation, which also demonstrated high accuracy and low variability (84$\pm$3\% and 79$\pm$8\%, respectively). However, their performance for artificial structures was notably weaker, with a significant disparity between the two indices (32$\pm$22\% for \gls{evi} and 64$\pm$23\% for \gls{evi2}). Among the two, \gls{evi2} achieved superior accuracy, making it the more effective index.

The \gls{savi} and \gls{osavi} indices demonstrated significantly better performance for vegetation (87$\pm$4\% and 89$\pm$4\%, respectively) compared to \gls{msavi} (54$\pm$1\%). However, the reverse was observed for water, where \gls{msavi} achieved the highest accuracy (99$\pm$1\%), slightly outperforming \gls{savi} and \gls{osavi} (both $\approx$97$\pm$2\%). For artificial structures, all three indices exhibited comparable accuracy (approximately 50\%); however, \gls{msavi} demonstrated much lower variability (4\%) compared to \gls{savi} and \gls{osavi} (both $\approx$25\%).

In terms of accuracy, \gls{mbi} and its enhanced version, \gls{embi}---which specialize in differentiating soils---exhibited similar performance for vegetation ($\approx$68\%) and artificial structures ($\approx$45\%). However, \gls{embi} demonstrated significantly higher accuracy for water (97\% compared to 78\% for \gls{mbi}). Additionally, \gls{embi} consistently showed markedly lower variability across all three features, highlighting its improved stability over the \gls{mbi}.

The \gls{tavi} index demonstrated moderate accuracy for water (88$\pm$16\%) and artificial structures (76$\pm$10\%), but performed significantly worse for vegetation (39$\pm$17\%). Notably, despite achieving higher accuracy in some cases, \gls{tavi} exhibited substantial variability across all three features, limiting its applicability in real-world scenarios. Designed to mitigate topographical reflectance effects, \gls{tavi} may have been disadvantaged in this study as the reflectance maps did not necessitate such corrections, potentially impacting its performance.

The water indices, \gls{ndwi} and \gls{mndwi}, performed expectedly well in water extraction, achieving accuracy values of 97$\pm$3\% and 99$\pm$1\%, respectively. However, their performance was weaker for vegetation (47$\pm$21\% and 33$\pm$21\%) and artificial structures (55$\pm$2\% and 63$\pm$7\%). In contrast, \gls{ndmi}, which specializes in detecting vegetation moisture, showed worse accuracy for water (73$\pm$8\%) and artificial structures (42$\pm$13\%), but achieved better results for vegetation (67$\pm$20\%). Despite this, \gls{ndmi} exhibited high variability across all three features, limiting its reliability.

The \gls{nbr} and \gls{bai} indices were primarily designed for extracting burnt areas. However, since the regions of interest in this study did not feature any recent fire activity, their evaluation for this particular characteristic was not possible. \gls{nbr} demonstrated mediocre to poor performance across all three features, with accuracy and variability values of 67$\pm$20\% for vegetation, 73$\pm$29\% for water, and 42$\pm$13\% for artificial structures. \gls{bai}, while performing poorly for water (3$\pm$2\%) and artificial structures (5$\pm$4\%), achieved very good results for vegetation (88$\pm$5\%). Its performance for vegetation was comparable to dedicated indices such as \gls{ndvi} (87$\pm$6\%), suggesting that \gls{bai} could be a viable alternative for vegetation extraction in certain contexts.

The \gls{asi} index displayed very good results for vegetation extraction (87$\pm$6\%) but performed poorly for artificial structures (13$\pm$17\%) and water (3$\pm$2\%), ranking among the lowest for these features. Its components—\gls{af}, \gls{vsf}, \gls{ssf}, and \gls{mf}—although not designed for direct use as standalone indices, yielded some notable results. \gls{af} achieved excellent accuracy for vegetation and water, but exhibited mediocre performance for artificial structures. Conversely, \gls{vsf} performed well for water, but showed poor results for vegetation. Both \gls{ssf} and \gls{mf} achieved good performance for vegetation but were less reliable for water due to high variability and for artificial structures due to low overall accuracy.

The road extraction indices, \gls{rei} and \gls{ri}, demonstrated mediocre to good performance for vegetation extraction (66$\pm$4\% and 88$\pm$5\%, respectively). However, their performance varied significantly for other features. For water extraction, \gls{ri} exhibited much lower accuracy (3$\pm$2\%) compared to \gls{rei} (88$\pm$12\%). Similarly, for artificial structure extraction, \gls{ri} showed significantly poorer accuracy (5$\pm$4\%) than \gls{rei} (38$\pm$6\%).

\section{Discussion and Recommendations}
\label{sub:recommendations}

It is important to recognize that factors such as regional vegetation composition, environmental conditions, and even the scale of analysis can contribute to the poor performance of certain indices, indicating that they may require further calibration or adjustment for optimal use. Nonetheless, the results obtained in Studies I and II, as well as the literature reviewed throughout this work, enable us to identify several indices that are particularly effective for specific feature extraction tasks. The main recommendations of this study are summarized in Table~\ref{tab:sum-index-rec}, which identifies the indices that consistently performed best for specific feature extraction tasks..

\begin{table}[ht]
\centering
\caption{Recommended spectral indices for major land-cover features based on their observed performance across the case studies and their documented effectiveness in the literature.}
\label{tab:sum-index-rec}
\begin{tabular}{lll}
\toprule
Feature & Index & Reference \\
\midrule
\multirow{5}{*}{Vegetation}
 & NDVI$^*$ & \cite{rouse-1974} \\
 & GNDVI & \cite{gitelson-1996} \\
 & GARI & \cite{kaufman-1992} \\
 & NDRE & \cite{tucker-1979} \\
 & BAI$^{\dagger}$ & \cite{chuvieco-2002} \\
\midrule
\multirow{1}{*}{Water}
 & MNDWI$^*$ & \cite{xu-2006} \\
\midrule
\multirow{1}{*}{Artificial surfaces}
 & MSR$^*$ & \cite{chen-1996} \\
\bottomrule
\multicolumn{3}{l}{\scriptsize $^*$ Top performer.}\\
\multicolumn{3}{l}{\scriptsize $^{\dagger}$ Effective indices not originally designed for the feature.}\\
\end{tabular}
\end{table}

The findings from our case studies on vegetation extraction yielded relatively consistent results, with several indices, such as \gls{ndvi}, \gls{gndvi}, \gls{gari}, and \gls{ndre}, reaffirming their effectiveness by demonstrating high average accuracies and low variability. Interestingly, some indices originally designed for other features, such as \gls{bai}, also proved effective for vegetation extraction. In this context, \gls{ndvi} stands as our recommended choice. This recommendation is based not only on \gls{ndvi}'s performance in the case studies but also on recent scientific literature supporting its efficacy and widespread use in vegetation cover assessment applications \citep{sidi2021spatiotemporal, cheng2022spatiotemporal}, particularly in segmentation tasks \citep{bosilj2018connected, ruan2020detecting}.

Water extraction was undoubtedly the most competitive category, with the majority of indices—regardless of their specific focus on water extraction—demonstrating exceptional results. This could suggest that the spectral reflectance characteristics of water bodies are more distinct and easier to isolate compared to other land cover types. However, it is important to consider that the relatively small area covered by water in the study region may have influenced these outcomes. In contrast, artificial structures, which also occupy a limited portion of the landscape, did not yield comparable results, suggesting that spatial extent may not be a significant factor contributing to the observed accuracy. Literature identifies \gls{mndwi} as a highly effective tool for water extraction \citep{masocha-2018}, a claim that is supported by the findings of Case Study II. \gls{mndwi} consistently achieved one of the highest accuracy ratings among the indices evaluated, coupled with low variability, further reinforcing its reliability in this context.

Finally, for artificial structure extraction, all the specialized indices---\gls{asi} and its components, \gls{rei}, and \gls{ri}---failed to yield satisfactory results. For \gls{asi}, the low accuracy and extreme variability observed in Study II raise concerns about the index's overall applicability. In contrast, for the road extraction indices, one could argue that they are primarily suited for road detection rather than artificial structures as a whole. Additionally, at the scale of analysis, the level of road detail may be too fine to capture meaningful results. This argument is further supported by the findings of Study I, where both indices struggled to delineate roads, even major thoroughfares, at a similar scale. While the remaining indices also demonstrate weak and generally unreliable results, \gls{msr} emerges as a strong choice for artificial structures, with high average accuracy and low variability, indicating its robust ability to differentiate between artificial structures and other features. However, the optimal threshold range for \gls{msr} in artificial structures is incredibly narrow across all features, meaning the decision boundary between the two classes is very tight. While this can result in high precision, as seen in the case, it also suggests that the index may be sensitive to noise, slight variations, or imperfections in the data. Consequently, it might not generalize well to data variations because it has to be finely tuned to such a narrow window.

\section{Constraints and Limitations}
\label{sub:limitations}

Several factors may limit the applicability of the methods presented in this study, including geographical and methodological constraints.

Firstly, the research was conducted within the context of mainland Portugal, which may restrict the direct applicability of the results to other regions. Nonetheless, the findings are likely relevant for areas with similar land cover and climate conditions, such as parts of Southern Europe---including Spain, Italy, and Greece---and the Northern Maghreb, such as Morocco, Algeria, and Tunisia.

Secondly, even with high-resolution land cover maps, the omission of small fires---including the early stages of large wildfires---remains a significant limitation. As noted by \citet{laneve2006continuous}, the under-detection of such events, often due to subpixel-scale fires, can distort estimates of fire frequency and spatial distribution. This leads to biased fire risk models and inefficient resource allocation. Moreover, the lack of representation of small fires can compromise long-term analyses by underestimating the effects of land use dynamics on fire regimes. Since small fires often serve as early indicators of larger events, their omission also reduces the effectiveness of early detection and response strategies.

Thirdly, given that burn indices could not be fully validated in this study, a robust validation framework is necessary for reliable burn index based monitoring. It is recommended to systematically compare burn area outputs from indices with reference data, including historical records, satellite-derived products, and field observations across different seasons and environments. Collaboration with fire management agencies is essential to obtain high-resolution ground truth data and establish standardized validation protocols. Field campaigns in fire-prone areas, combining on-site monitoring with spectral data collection, are encouraged to refine thresholds and contextualize index results. The use of multi-sensor data---such as \gls{sar} fusion and thermal anomaly detection \citep{laneve2024application}---is also suggested to improve burn area mapping, especially under conditions of cloud cover, smoke, or complex terrain.

Lastly, the quality of land cover maps significantly affects the accuracy of burn area products, yet standardization remains limited. Variations in spatial resolution, classification schemes, and update frequency across land cover datasets can lead to inconsistencies in burn area estimates, particularly in dynamic landscapes. These discrepancies may reduce spatial precision and introduce bias in trend and risk assessments. To improve comparability and reliability, the use of standardized land cover data is recommended, or at least a clear report of the datasets and classification methods used should be provided. This transparency supports reproducibility and helps clarify how land cover influences burn area detection.

\section{Conclusions}
\label{sub:conclusions}

This article presented a review of existing literature on multispectral indices and their applications in extracting specific landscape features, such as vegetation, soil, water, artificial structures, and burnt areas. Additionally, two case studies were presented to assess the effectiveness of key indices selected from the literature, focusing on the extraction of vegetation, water, and artificial structures from satellite imagery.

In the first case study, a qualitative review was conducted to evaluate eight of the most significant indices for various extraction purposes. Their performance was analyzed by examining the output images and considering the foundational principles underlying each index as discussed in the literature. In the second study, a more objective approach was adopted by directly comparing the segmentation outputs of 33 multispectral indices with an official ground-truth land use map. This method enabled the determination of an accuracy rating for each index. Results lead to the identification of the most effective indices for extracting specific landscape features in the regions under analysis. For vegetation extraction, \gls{ndvi} reaffirmed its position as a reliable choice, demonstrating robust performance consistent with extensive validation in the scientific literature. Regarding water extraction, \gls{mndwi} was confirmed as the most consistent and accurate tool for identifying water features. Finally, \gls{msr} emerged as the most effective index for extracting artificial structures, demonstrating high accuracy.

Building on these insights, several recommendations for future research can be proposed. A key priority should be gaining a deeper understanding of the regional specificity of each index in relation to particular landscape features. This understanding would enable the more effective use of supporting data, allowing these indices to reach their full potential. Integrating advanced technologies, such as machine learning algorithms and data fusion techniques, holds significant potential to improve the accuracy and reliability of multispectral indices across diverse applications, including those that have shown suboptimal performance \citep{hoffren2023uav, iheaturu2024integrating}. Future studies should explore these approaches to optimize extraction and analysis processes, enabling the generation of more precise and actionable data, especially for critical applications such as wildfire management.

Despite recent advances in machine learning, traditional multispectral indices remain crucial for effective and interpretable burn area detection and assessment. While modern machine learning techniques can offer higher accuracy in complex classification tasks, they often require large labeled datasets, significant computational resources, and may lack transparency. In contrast, traditional indices are well-established, easy to compute, and provide reliable baseline information. Integrating these indices as input features or for pre-classification can enhance model accuracy, reduce training time, and support interpretation by grounding results in domain knowledge. Therefore, traditional multispectral indices continue to play a valuable complementary role in machine learning–based fire mapping workflows.

Lastly, while hyperspectral imaging represents a promising frontier in remote sensing for wildfire management, multispectral imaging currently provides the optimal balance of cost-effectiveness and ease of use. Continued research should focus on refining multispectral indices and exploring the integration of hyperspectral data to advance wildfire management strategies, paving the way for more sophisticated and precise monitoring and management solutions.

\section*{Funding}

{ 
This research was partially funded by the Funda\c c\~ao para a Ci\^encia e a
Tecnologia (FCT, \url{https://ror.org/00snfqn58}) under grants/projects
UID\-/\-6486\-/\-2025
(\url{https://doi.org/10.54499/UID/06486/2025}),
UID\-/\-PRR\-/\-6486\-/\-2025
(\url{https://doi.org/10.54499/UID/PRR/06486/2025}),
UID\-/\-PRR2\-/\-06486\-/\-2025
(\url{https://doi.org/10.54499/UID/PRR2/06486/2025}),
2023.15441.TENURE.051\-/\-CP00003\-/\-CT00029,
UID\-/\-00408\-/\-2025
(\url{https://doi.org/10.54499/UID/00408/2025}),
UID\-/\-PRR\-/\-00408\-/\-2025
(\url{https://doi.org/10.54499/UID/PRR/00408/2025}),
UID\-/\-00066\-/\-2025
(\url{https://doi.org/10.54499/UID/00066/2025}),
UID\-/\-PRR\-/\-00066\-/\-2025
(\url{https://doi.org/10.54499/UID/PRR/00066/2025}),
and CEECINST\-/\-00002\-/\-2021\-/\-CP2788\-/\-CT0001
(\url{https://doi.org/10.54499/CEECINST/00002/2021/CP2788/CT0001}),
as well as by the Instituto Lusófono de Investigação e Desenvolvimento (ILIND, \url{https://ror.org/02qy8ba98}), Portugal, under Project COFAC\-/\-ILIND\-/\-COPELABS\-/\-1\-/\-2024
(\url{ https://doi.org/10.62658/COFAC/ILIND/COPELABS/1/2024}).
}

\section*{Conflict of Interest Statement}

The authors declare that the research was conducted in the absence of any commercial or financial relationships that could be construed as a potential conflict of interest.

\section*{Data Availability Statement}
The data analysed during this study is available upon request.

\section*{Generative AI Statement}
The author(s) declared that generative AI was used in the creation of this manuscript. The authors used ChatGPT for language editing purposes, including spell checking and improving the flow of selected passages of the manuscript. All ideas, results, analyses, and conclusions presented in this paper are original and were produced under the full responsibility of the authors, who have reviewed and approved the final version of the manuscript.

\appendix

\bibliographystyle{elsarticle-harv}

\appendix

\clearpage
\section{Index Acronym Table}
\label{sub:annex-acronyms}

\begin{table}[!htpb]
\begin{center}
\caption{Multispectral indices acronym reference table.\label{tab:acronym-indices}}
\scriptsize
\begin{tabular}{ll}
    \toprule
    Acronym & Name \\
    \cmidrule(rl){1-1}\cmidrule(rl){2-2}
    ABAI & Analytic Burned Area Index \\
    ARVI & Atmospherically Resistant Vegetation Index \\
    ARI & Anthocyanin Reflectance Index \\
    ASI & Artificial Surface Index \\
    BAI & Burning Area Index \\
    BAIS2 & Burned Area Index for Sentinel 2 \\
    BCI & Biophysical Composition Index \\
    BLFEI & Built-Up Land Features Extraction Index \\
    EMBI & Enhanced Modified Bare Soil Index \\
    DVI & Difference Vegetation Index \\
    EVI & Enhanced Vegetation Index \\
    EVI2 & Enhanced Vegetation Index 2 \\
    FVI & Fusion Vegetation Index \\
    GNDVI & Green Normalized Difference Vegetation Index \\
    HEVI2 & Hotspot-Signature 2-Band Enhanced Vegetation Index \\
    HSVI & Hotspot-Signature Soil-adjusted Vegetation Index \\
    IBI & Index-Based Built-Up Index \\
    MARI & Modified Anthocyanin Reflectance Index \\
    MBI & Modified Bare Soil Index \\
    MCARI & Modified Chlorophyll Absorption in Reflectance Index \\
    MNDWI & Modified Normalized Difference Water Index \\
    MSR & Modified Simple Ratio Index \\
    MuWi & Multi-Spectral Water Index \\
    MSAVI & Modified Soil-Adjusted Vegetation Index \\
    MTVI2 & Modified Triangular Vegetation Index \\
    NBR & Normalized Burn Ratio Index \\
    NBRT1 & Normalized Burn Ratio Thermal Index \\
    NDBI & Normalized Difference Built-Up Index \\
    NDHD & Normalized Difference Between Hotspot and Darkspot \\
    NDMI & Normalized Difference Moisture Index \\
    NDRE & Normalized Difference Red Edge Index \\
    NDVI & Normalized Difference Vegetation Index \\
    NDWI & Normalized Difference Water Index \\
    OSAVI & Optimized Soil-Adjusted Vegetation Index \\
    PISI & Perpendicular Impervious Surface Index \\
    PRI & Photochemical Reflectance Index \\
    RDVI & Renormalized Difference Vegetation Index \\
    REI & Road Extraction Index \\
    REVI & Red-Edge Vegetation Index \\
    RI & Road Index \\
    RVI & Radio Vegetation Index \\
    SAVI & Soil-Adjusted Vegetation Index \\
    SR & Simple Ratio Index \\
    TCARI & Transformed Chlorophyll Absorption in Reflectance Index \\
    TAVI & Topography-Adjusted Vegetation Index \\
    UI & Urban Index \\
    UPDM & Universal Pattern Decomposition Method \\
    VIUPD & Vegetation Index Based on Universal Pattern Decomposition Method \\
    VrNIR-BI & Visible Red and NIR-Based Built-Up Index \\
    \bottomrule
    \end{tabular}
\end{center}
\end{table}

\clearpage
\section{Supplementary Index Table}
\label{sub:annex-formulas}

\begin{table}[!htpb]
\begin{center}
\caption{Supplementary multispectral index table.\label{tab:sup-indices}}
\footnotesize
\begin{tabular}{llc}
    \toprule
    Index & Formula & Ref. \\
    \midrule
    UI
        & $\frac{\rho_{\text{SWIR2}}-\rho_{\text{NIR}}}{\rho_{\text{SWIR2}}+\rho_{\text{NIR}}}$
        & \citep{kawamura-1996}  \\
    NDBI
        & $\frac{\rho_{\text{SWIR1}}-\rho_{\text{NIR}}}{\rho_{\text{SWIR1}}+\rho_{\text{NIR}}}$
        & \citep{zha-2003}  \\
    IBI
        & $\frac{\rho_{\text{NDBI}}-(\rho_{\text{SAVI}}+\rho_{\text{MNDWI}})/2}{\rho_{\text{NDBI}}+(\rho_{\text{SAVI}}+\rho_{\text{MNDWI}})/2}$
        & \citep{zha-2003}  \\
    BCI
        & $\frac{(\text{H}+\text{L})/2-\text{V}}{(\text{H}+\text{L})/2+\text{V}}$
        & \citep{deng-2012}  \\
    VgNIR-BI
        & $\frac{\rho_{\text{G}}-\rho_{\text{NIR}}}{\rho_{\text{G}}+\rho_{\text{NIR}}}$
        & \citep{estoque-2015}  \\
    PISI
        & $0.8192\times \rho_{\text{B}}-0.5735\times \rho_{\text{NIR}}+0.0750$
        & \citep{tian-2018}  \\
    BLFEI
        & $\frac{(\rho_{\text{G}}+\rho_{\text{R}}+\rho_{\text{SWIR2}})/3-\rho_{\text{SWIR1}}}{(\rho_{\text{G}}+\rho_{\text{R}}+\rho_{\text{SWIR2}})/3+\rho_{\text{SWIR1}}}$
        & \citep{bouhennache-2019}  \\
    MuWi
        & $ \sum_{}^{}a_{i_{\rho_{i}}} - b$
        & \citep{bouhennache-2019}  \\
    BAIS2
        & $\left(1-\sqrt{\frac{\rho_{\text{RE}_{06}}\times\rho_{\text{RE}_{07}}\times\rho_{\text{RE}_{8\text{A}}}}{\rho_{\text{R}}}}\right) \times \left(\frac{\rho_{\text{SWIR2}}-\rho_{\text{RE}_{8\text{A}}}}{\sqrt{\rho_{\text{SWIR2}}+\rho_{\text{RE}_{8\text{A}}}}}+1\right)$
        & \citep{filipponi2018bais2} \\
    ABAI
        & $\frac{3\times\rho_{\text{SWIR2}}-2\times\rho_{\text{SWIR1}}-3\times\rho_{\text{G}}}{\rho_{3\times\text{SWIR2}}+2\times\rho_{\text{SWIR1}}+3\times\rho_{\text{G}}}$
        & \citep{wu2022forest} \\
    \bottomrule
    & & \\
    \multicolumn{3}{l}{\scriptsize $\qquad\rho_{\text{B}}$, $\rho_{\text{G}}$, $\rho_{\text{R}}$, $\rho_{\text{RE}}$, $\rho_{\text{NIR}}$, $\rho_{\text{NNIR}}$, $\rho_{\text{SWIR1}}$, $\rho_{\text{SWIR2}}$: Reflectance for the Blue, Green, Red,}\\
    \multicolumn{3}{l}{\scriptsize Red Edge, NIR, Narrow NIR, and SWIR band reflectances, respectively.}\\
    \multicolumn{3}{l}{\scriptsize \qquad\text{H}, \text{V}, \text{L}: Normalized brightness, greenness, and wetness components from the}\\
    \multicolumn{3}{l}{\scriptsize Tasseled Cap (TC) transformation.}\\

    \end{tabular}
\end{center}
\end{table}

\clearpage
\section{Supplementary Practical Case Study I Figures}
\label{sub:annex-figures}

\begin{figure}[!htpb]
\begin{center}
\footnotesize
\includegraphics[width=.75\textwidth]{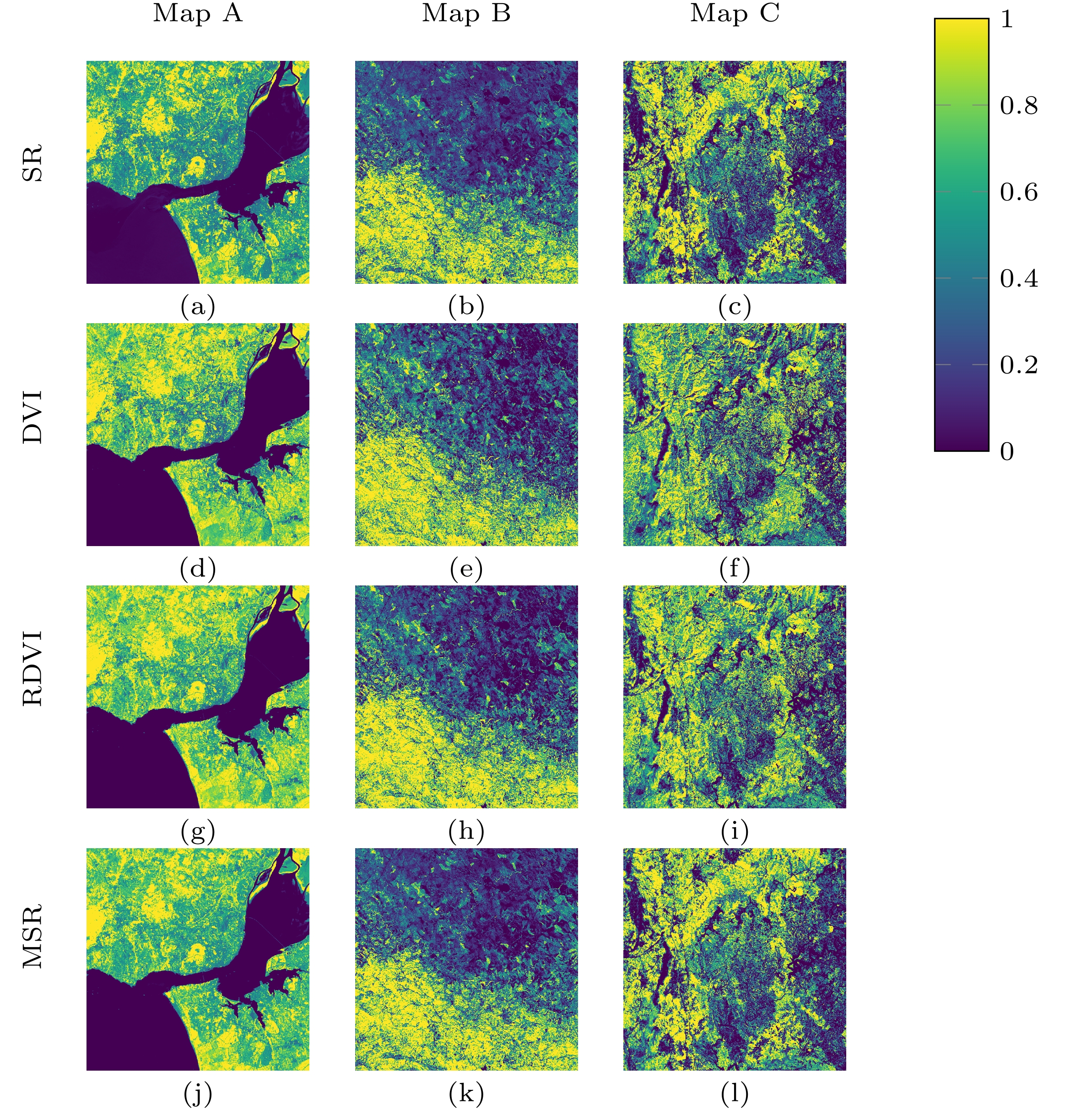}
\caption{Case Study I results for the simple greenness indicators (I). Columns A, B, and C represent three distinct test regions: the greater Lisbon area, southern Portugal, and central Portugal, respectively. Every row showcases the outcomes of distinct indices: SR, DVI, RDVI, and MSR. All indices are visualized using the \textit{Viridis} colormap, with values normalized according to the provided scale bar.\label{tab:practical-simple-1}}
\end{center}
\end{figure}

\begin{figure}[!htpb]
\begin{center}
\footnotesize
\includegraphics[width=.75\textwidth]{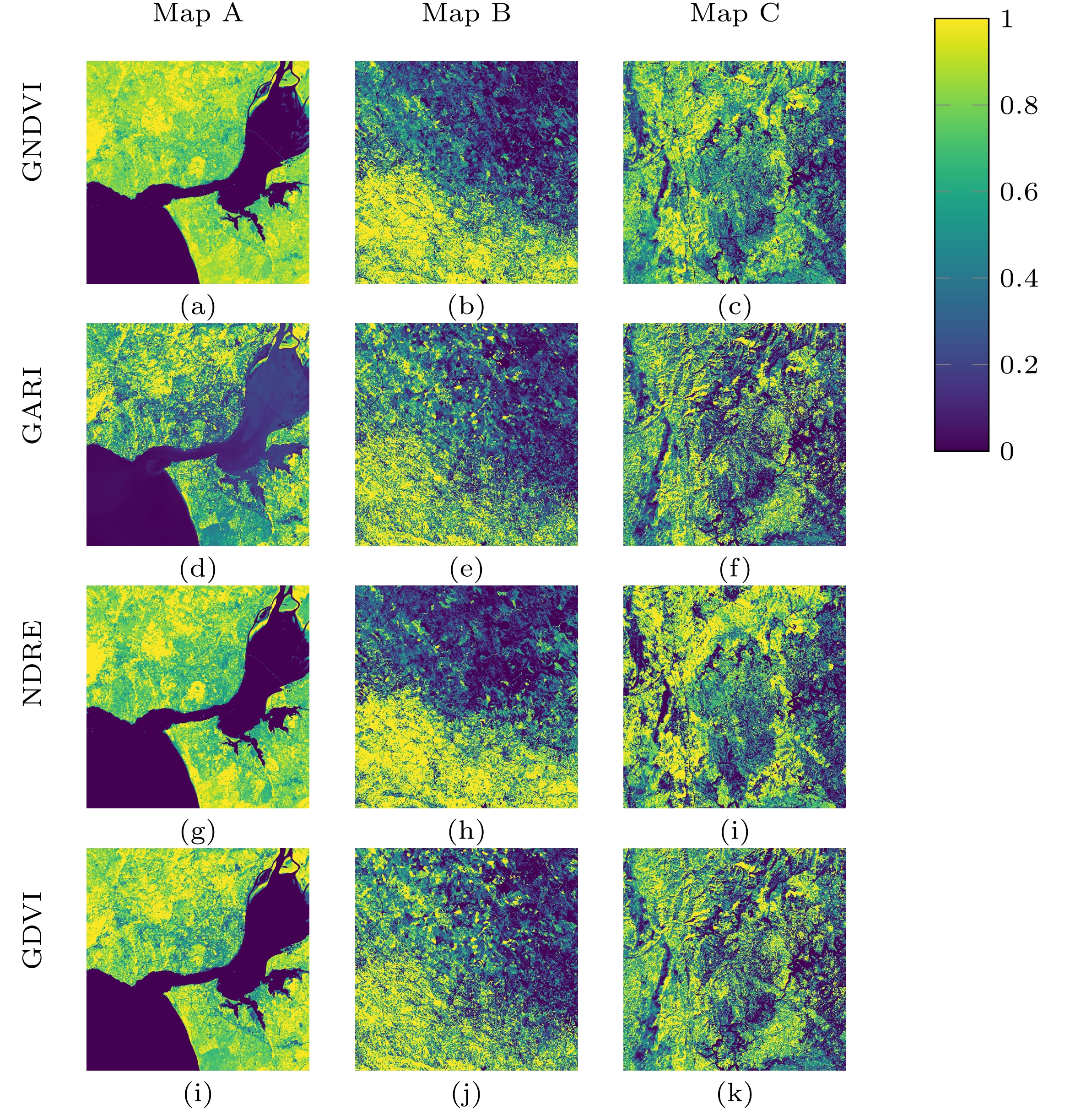}
\caption{Case Study I results for the simple greenness indicators (II). Columns A, B, and C represent three distinct test regions: the greater Lisbon area, southern Portugal, and central Portugal, respectively. Every row showcases the outcomes of distinct indices: GNDVI, GARI, NDRE, and GDVI. All indices are visualized using the \textit{Viridis} colormap, with values normalized according to the provided scale bar.\label{tab:practical-simple-2}}
\end{center}
\end{figure}

\begin{figure}[!htpb]
\begin{center}
\footnotesize
\includegraphics[width=.75\textwidth]{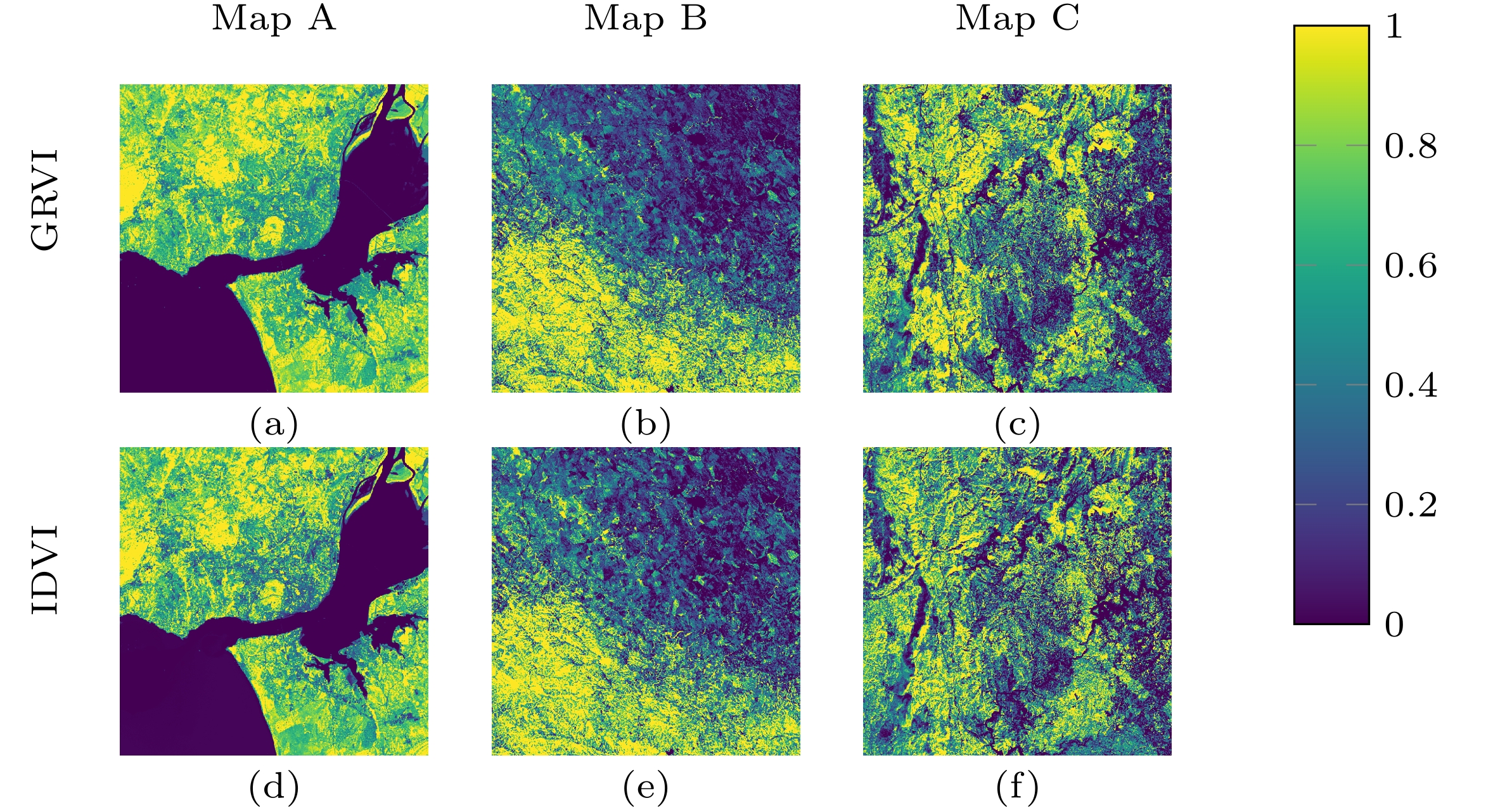}
\caption{Case Study I results for the simple greenness indicators (III). Columns A, B, and C represent three distinct test regions: the greater Lisbon area, southern Portugal, and central Portugal, respectively. Every row showcases the outcomes of distinct indices: GRVI and IDVI. All indices are visualized using the \textit{Viridis} colormap, with values normalized according to the provided scale bar.\label{tab:practical-simple-3}}
\end{center}
\end{figure}

\begin{figure}[!htpb]
\begin{center}
\footnotesize
\includegraphics[width=.75\textwidth]{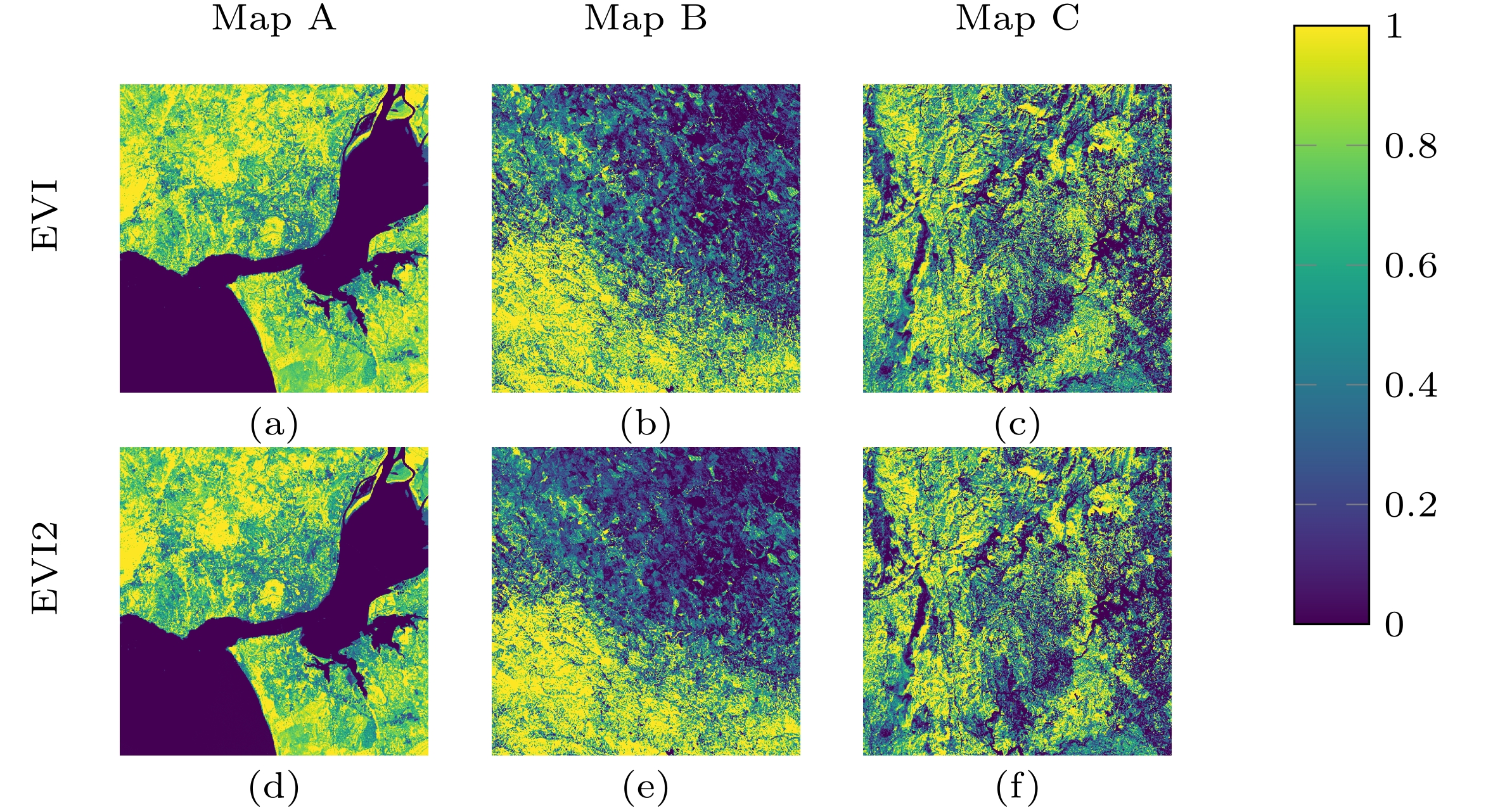}
\caption{Case Study I results for the enhanced vegetation indices. Columns A, B, and C represent three distinct test regions: the greater Lisbon area, southern Portugal, and central Portugal, respectively. Every row showcases the outcomes of distinct indices: EVI and EVI2. All indices are visualized using the \textit{Viridis} colormap, with values normalized according to the provided scale bar.\label{tab:practical-enhanced}}
\end{center}
\end{figure}

\begin{figure}[!htpb]
\begin{center}
\footnotesize
\includegraphics[width=.75\textwidth]{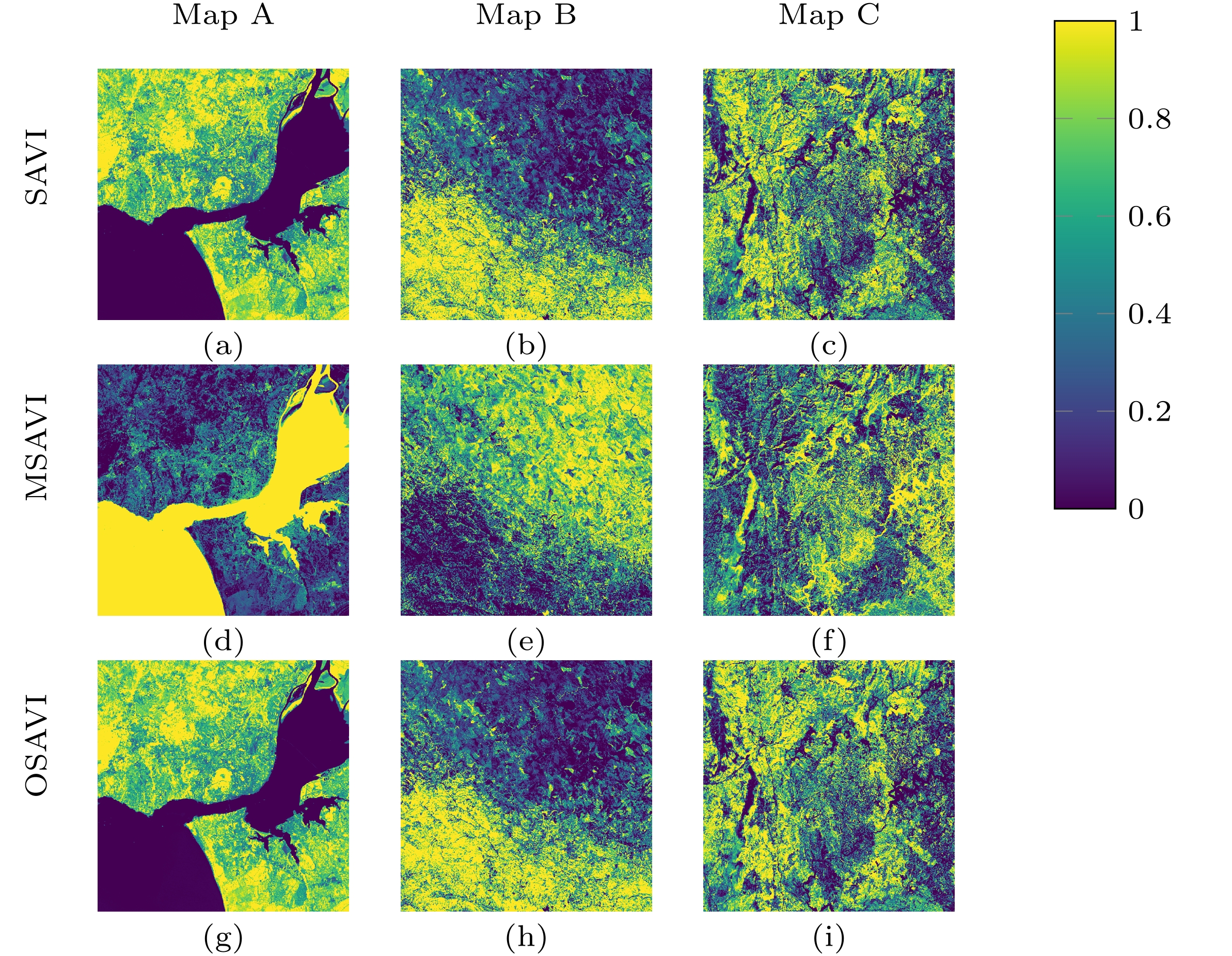}
\caption{Case Study I results for the soil-adjusted vegetation indices. Columns A, B, and C represent three distinct test regions: the greater Lisbon area, southern Portugal, and central Portugal, respectively. Every row showcases the outcomes of distinct indices: SAVI, MSAVI, and OSAVI. All indices are visualized using the \textit{Viridis} colormap, with values normalized according to the provided scale bar.\label{tab:practical-sa}}
\end{center}
\end{figure}

\begin{figure}[!htpb]
\begin{center}
\footnotesize
\includegraphics[width=.75\textwidth]{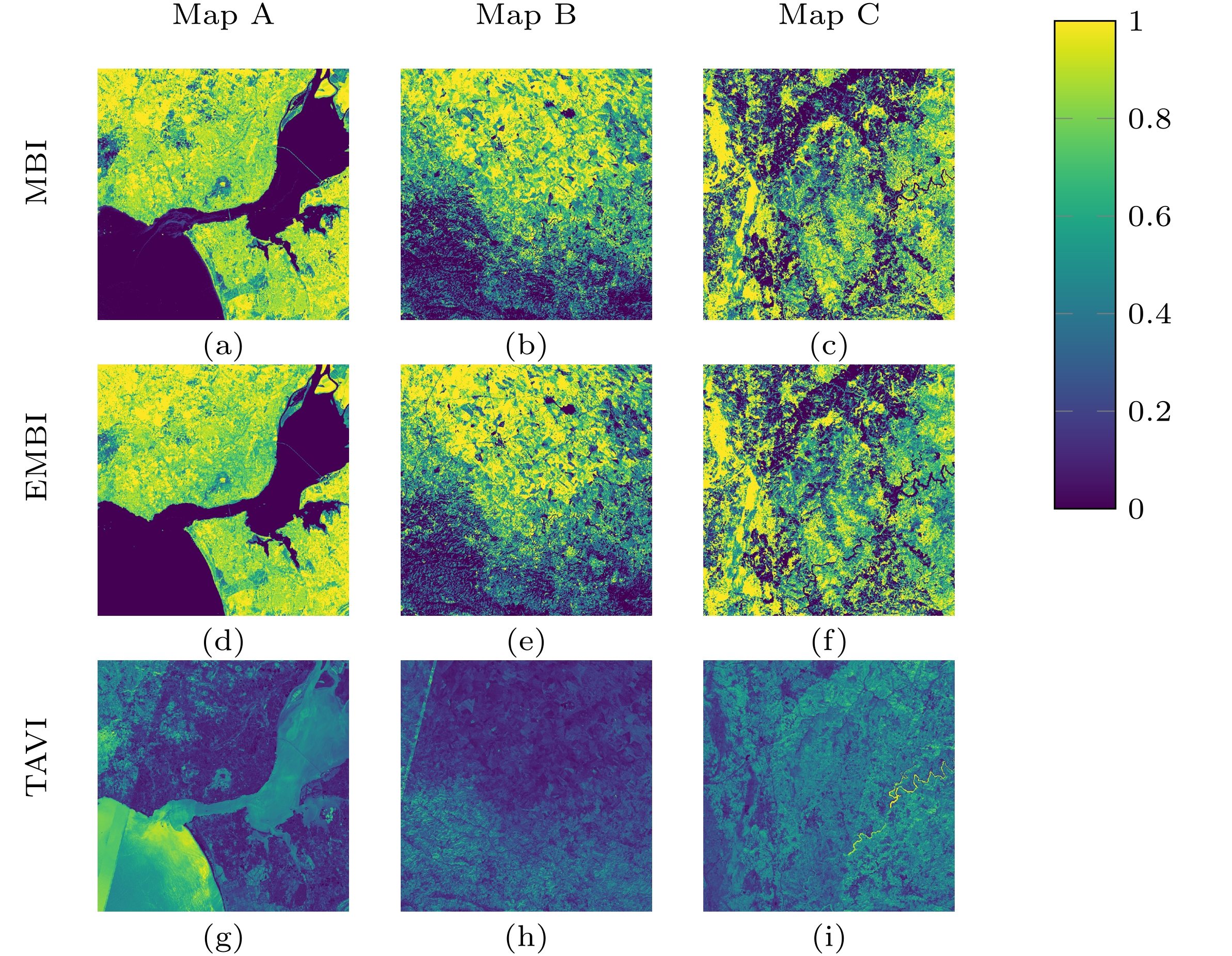}
\caption{Case Study I results for the modified bare soil indices and the \acrshort{tavi}. Columns A, B, and C represent three distinct test regions: the greater Lisbon area, southern Portugal, and central Portugal, respectively. Every row showcases the outcomes of distinct indices: MBI, EMBI, and TAVI. All indices are visualized using the \textit{Viridis} colormap, with values normalized according to the provided scale bar.\label{tab:practical-bare-tavi}}
\end{center}
\end{figure}

\begin{figure}[!htpb]
\begin{center}
\footnotesize
\includegraphics[width=.75\textwidth]{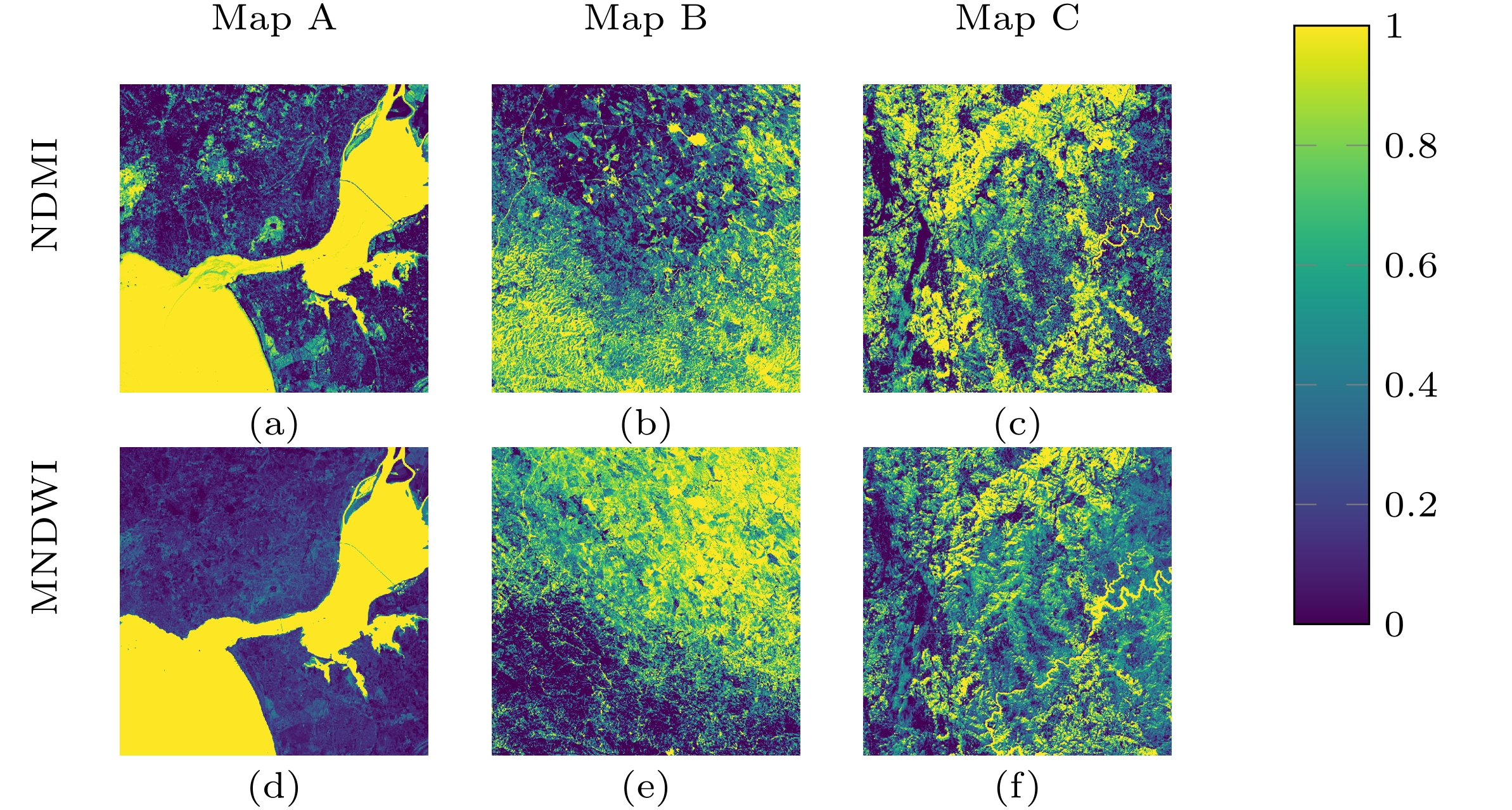}
\caption{Case Study I results for the water/moisture extraction indices. Columns A, B, and C represent three distinct test regions: the greater Lisbon area, southern Portugal, and central Portugal, respectively. Every row showcases the outcomes of distinct indices: NDMI and MNDWI. All indices are visualized using the \textit{Viridis} colormap, with values normalized according to the provided scale bar.\label{tab:practical-water}}
\end{center}
\end{figure}

\begin{figure}[!htpb]
\begin{center}
\footnotesize
\includegraphics[width=.75\textwidth]{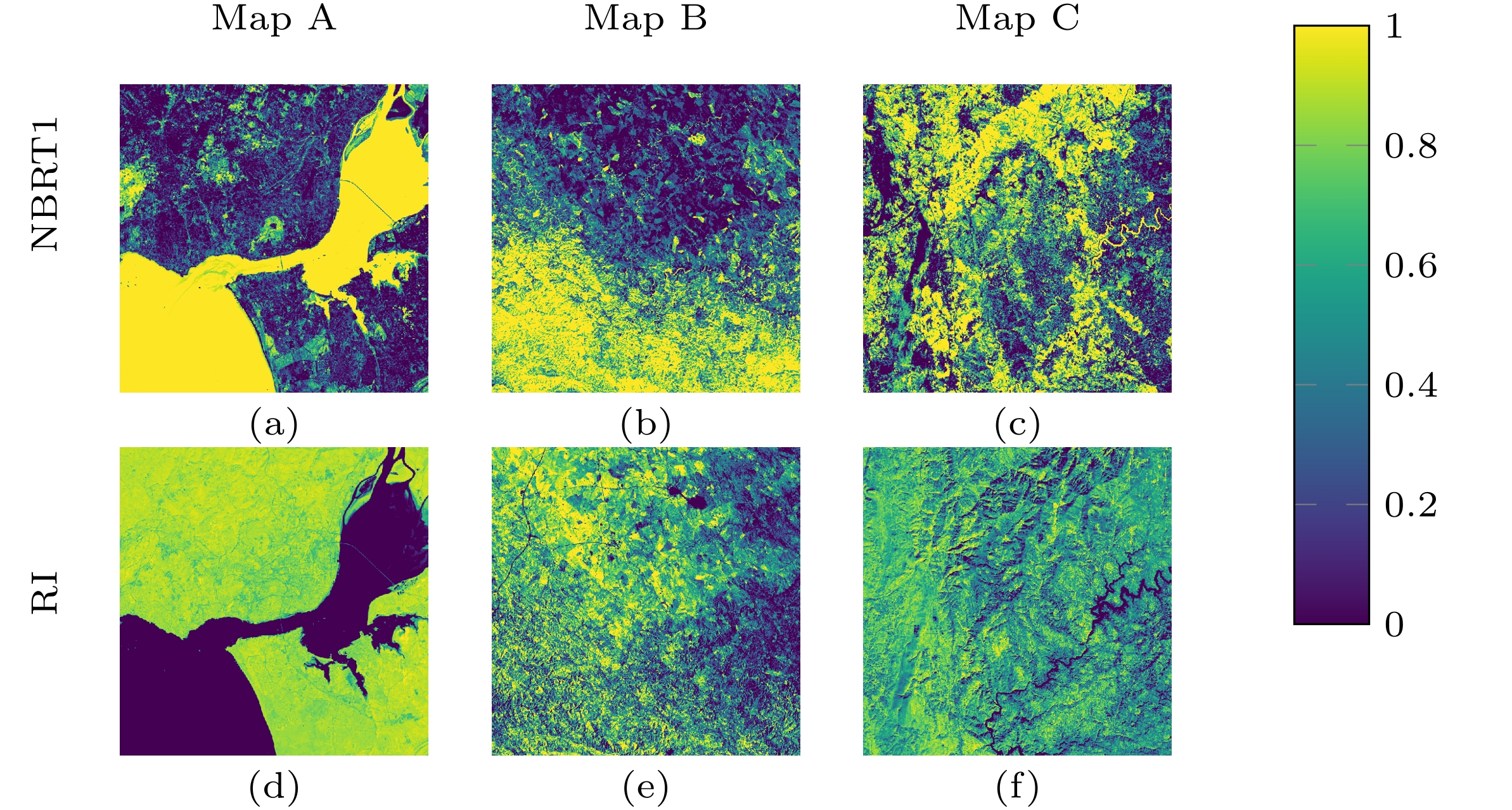}
\caption{Practical study results for the \acrshort{nbrt1} and \acrshort{ri}. Columns A, B, and C represent three distinct test regions: the greater Lisbon area, southern Portugal, and central Portugal, respectively. Every row showcases the outcomes of distinct indices: NBRT1 and RI. All indices are visualized using the \textit{Viridis} colormap, with values normalized according to the provided scale bar.\label{tab:practical-burnt}}
\end{center}
\end{figure}

\end{document}